\PassOptionsToPackage{unicode}{hyperref}
\PassOptionsToPackage{hyphens}{url}
\documentclass[
  11pt,
]{article}
\usepackage{lmodern}
\usepackage{amssymb,amsmath}
\usepackage{ifxetex,ifluatex}
\ifnum 0\ifxetex 1\fi\ifluatex 1\fi=0 
  \usepackage[T1]{fontenc}
  \usepackage[utf8]{inputenc}
  \usepackage{textcomp} 
\else 
  \usepackage{unicode-math}
  \defaultfontfeatures{Scale=MatchLowercase}
  \defaultfontfeatures[\rmfamily]{Ligatures=TeX,Scale=1}
\fi
\IfFileExists{upquote.sty}{\usepackage{upquote}}{}
\IfFileExists{microtype.sty}{
  \usepackage[]{microtype}
  \UseMicrotypeSet[protrusion]{basicmath} 
}{}
\makeatletter
\@ifundefined{KOMAClassName}{
  \IfFileExists{parskip.sty}{%
    \usepackage{parskip}
  }{
    \setlength{\parindent}{0pt}
    \setlength{\parskip}{6pt plus 2pt minus 1pt}}
}{
  \KOMAoptions{parskip=half}}
\makeatother
\usepackage{xcolor}
\IfFileExists{xurl.sty}{\usepackage{xurl}}{} 
\IfFileExists{bookmark.sty}{\usepackage{bookmark}}{\usepackage{hyperref}}
\hypersetup{
  pdftitle={Arithmetic Variable LogLog: Advancing the Memory-Variance Frontier},
  pdfauthor={Brian Bushnell1*},
  hidelinks,
  pdfcreator={LaTeX via pandoc}}
\usepackage[margin=0.5in]{geometry}
\usepackage{longtable,booktabs}
\usepackage{etoolbox}
\makeatletter
\patchcmd\longtable{\par}{\if@noskipsec\mbox{}\fi\par}{}{}
\makeatother
\IfFileExists{footnotehyper.sty}{\usepackage{footnotehyper}}{\usepackage{footnote}}
\makesavenoteenv{longtable}
\usepackage{graphicx}
\makeatletter
\def\maxwidth{\ifdim\Gin@nat@width>\linewidth\linewidth\else\Gin@nat@width\fi}
\def\maxheight{\ifdim\Gin@nat@height>\textheight\textheight\else\Gin@nat@height\fi}
\makeatother
\setkeys{Gin}{width=\maxwidth,height=\maxheight,keepaspectratio}
\makeatletter
\def\fps@figure{htbp}
\makeatother
\providecommand{\tightlist}{%
  \setlength{\itemsep}{0pt}\setlength{\parskip}{0pt}}
\newlength{\cslhangindent}
\newenvironment{cslreferences}%
  {}%
  {\par}

\title{Arithmetic Variable LogLog: Advancing the Memory-Variance
Frontier}
\author{Brian Bushnell\textsuperscript{1}*}
\date{2026}

\begin{document}
\maketitle

\textsuperscript{1}DOE Joint Genome Institute, Lawrence Berkeley
National Laboratory, Berkeley, CA, USA

*Corresponding author: bbushnell@lbl.gov

ORCiD: Brian Bushnell: https://orcid.org/0000-0002-8140-0131

\hypertarget{abstract}{%
\section{Abstract}\label{abstract}}

Cardinality estimation --- counting the number of distinct elements in a
data stream --- requires a tradeoff between memory and accuracy.
ExaLogLog {[}1{]} recently established the state of the art for this
tradeoff by combining wide registers with a Fisher-information-optimal
maximum likelihood (ML) estimator, achieving the best known
memory-variance product (MVP) among HyperLogLog {[}2{]} variants.

Here we present Arithmetic Variable LogLog (AVLL), which surpasses
ExaLogLog at every memory point tested using arithmetic encoding and
eliminating uncommon states to consume 64-bit words completely with 11
registers each, yielding a 5.5$\times$ register-count advantage. Its
four-component blended estimator, HLDLC, exploits this density advantage
to surpass ExaLogLog's ML accuracy without iterative solving.

At 1 KB, AVLL achieves 1.63\% width-weighted mean absolute error
compared to ExaLogLog's 1.71\% --- a 4.7\% improvement. The
corresponding empirical MVP is \(\approx\) 3.4, surpassing ExaLogLog's
practical MVP of 3.78 and its theoretical optimum of 3.67. This holds at
every tested size from 0.25 to 4 KB. AVLL inherits DynamicLogLog's early
exit mechanism, which filters most elements before any register is
touched. With thousands of simultaneous sketches per thread, AVLL is
2.7--4.5$\times$ faster than ExaLogLog due to the reduced memory bandwidth from
early exits.

Like DynamicLogLog, AVLL stores relative NLZ values with a shared
offset, so its memory scales as O(\emph{B} + log log \emph{C}) rather
than O(\emph{B} $\times$ log log \emph{C}) --- decoupling maximum representable
cardinality from register width. These results hold under both
high-complexity (all-unique) and low-complexity (nonuniformly high
duplication rate) data distributions, with zero accuracy degradation
from duplication. AVLL is implemented as a single self-contained Java
class with all correction formulas embedded, available in the BBTools
suite {[}3{]} at \url{https://bbmap.org}.

\hypertarget{introduction}{%
\section{1. Introduction}\label{introduction}}

The cardinality estimation problem --- counting distinct elements in a
data stream using bounded memory --- has been studied for four decades
since Flajolet and Martin's foundational work {[}4{]}. The practical
gold standard remains HyperLogLog (HLL) {[}2{]}, which achieves roughly
\(1.04/\sqrt{B}\) standard error using \emph{B} registers of 6 bits
each. In a packed implementation, 2,560 registers fit in 2 KB, yielding
approximately 1.65\% width-weighted mean absolute error.

Recent advances have pushed the memory-variance frontier in two
directions:

\textbf{Richer per-register information.} UltraLogLog (ULL) {[}5{]} adds
2-bit sub-NLZ history per register, enabling the FGRA (Further
Generalized Remaining Area) estimator to extract more information per
register and reduce the memory-variance product by 28\% relative to HLL.
ExaLogLog {[}1{]} extends this approach with variable-width registers
(6+\emph{t}+\emph{d} bits) and a Fisher-information-optimal maximum
likelihood estimator, achieving the best known memory-variance product
among fixed-register HLL variants.

\textbf{Architectural compaction.} DynamicLogLog (DLL) {[}6{]} uses a
shared exponent across all registers, reducing storage to 4 bits per
register --- a 33\% savings over HLL. At equal memory, DLL's extra
registers reduce variance by \(\sqrt{3/2} \approx 1.22\times\). DLL also
introduces Dynamic Linear Counting (DLC), a tier-aware extension of
Linear Counting that eliminates HLL's characteristic error spike at the
LC-to-LogLog transition.

These two approaches are orthogonal: one increases information density
per register, the other increases register density per byte. The fusion
of DLL and ULL (UDLL, 6 bits = 4 NLZ + 2 history per register)
demonstrated their compatibility, achieving ULL-level accuracy at 75\%
of the memory {[}6{]}.

AVLL's central observation is that once history is included, many states
become exceedingly rare and can be ignored with minimal loss of
accuracy, while fractional bits can be allocated to increase the number
of useful states per register. A LogLog register that distinguishes 56
states --- an NLZ exponent plus two bits of sub-NLZ history --- does not
need a 6-bit slot. Treating each 64-bit word as an 11-digit base-56
number packs eleven such registers per word with zero wasted bits: 5.82
bits per register, versus 6 bits and 8 wasted states for fixed-width
packing. At equal memory this compounds into a 5.5$\times$ register-count
advantage over ExaLogLog.

\textbf{Contributions.}

\begin{enumerate}
\def\labelenumi{\arabic{enumi}.}
\item
  \textbf{Arithmetic register encoding.} We demonstrate that base-56
  arithmetic encoding of LogLog registers is practical, yielding 11
  registers per 64-bit word with zero wasted bits. The encoding supports
  2-bit sub-NLZ history per register, tier promotion, and register-wise
  merge operations.
\item
  \textbf{HLDLC estimator.} A four-component blended estimator combining
  two complementary frameworks: LDLC (a DLC + HC blend, LC-based) and
  Hybrid+2 (a SBS + Mean+H blend, harmonic-mean-based). DLC and HC are
  inherited from {[}6{]}; Mean+H is a novel history-corrected harmonic
  mean; SBS is fully described here for the first time.
\item
  \textbf{Record accuracy per bit.} At every tested memory point
  (0.25--4 KB), AVLL achieves 4--5\% lower width-weighted error than
  ExaLogLog, with an empirical memory-variance product of approximately
  3.4 --- the lowest yet demonstrated, below ExaLogLog's theoretical
  optimum of 3.67.
\item
  \textbf{Deployment-scale speed.} Under cache pressure (thousands of
  simultaneous sketches per thread), AVLL's early exit mechanism yields
  2.7--4.5$\times$ ExaLogLog's throughput because most elements are filtered
  without touching register memory (Table 11).
\item
  \textbf{Unbounded cardinality at near-constant memory.} AVLL's
  sliding-floor architecture makes the cardinality contribution to
  memory additive rather than multiplicative: O(\emph{B} + log log
  \emph{C}) total, with no register-width-imposed ceiling.
\item
  \textbf{Comprehensive comparison.} We compare AVLL against seven other
  cardinality estimators --- ExaLogLog, HLL-TailCut4, HyperLogLogLog,
  Apache HLL4, HyperTwoBits, UltraLogLog, and LogLog30 --- under both
  high-complexity and low-complexity data distributions, demonstrating
  that AVLL is the most accurate at every tested memory point.
\item
  \textbf{Self-contained implementation.} A single Java class
  (ArithmeticVariableLogLog.java, \textasciitilde1,000 lines) with all
  correction formulas embedded as approximately 100 closed-form
  constants --- no external tables, no runtime file dependencies. AVLL
  is distributed with the BBTools suite {[}3{]} and is usable for
  \emph{k}-mer cardinality counting via \texttt{loglog.sh} with the flag
  \texttt{loglogtype=avll}.
\end{enumerate}

The remainder of this paper is organized as follows. Section 2 reviews
the relevant background. Section 3 describes AVLL's arithmetic encoding,
register architecture, and cardinality-scaling properties. Section 4
presents the HLDLC estimation pipeline and its four component
estimators. Section 5 describes the simulation methodology. Section 6
presents experimental results. Section 7 discusses the results and
limitations. Section 8 concludes.

\hypertarget{background}{%
\section{2. Background}\label{background}}

\hypertarget{the-loglog-framework}{%
\subsection{2.1 The LogLog Framework}\label{the-loglog-framework}}

All LogLog-family estimators share a common structure: a hash function
maps elements to uniformly distributed integers, the hash is split into
a bucket selector (determining which of \emph{B} registers to update)
and a rank (the number of leading zeros, NLZ, in the remaining bits),
and each register stores the maximum NLZ observed. The maximum NLZ
encodes information about the number of distinct elements: if \emph{n}
distinct elements are distributed across \emph{B} buckets, the expected
maximum NLZ per bucket is approximately \(\log_2\)(\emph{n}/\emph{B}).

We use the notation and definitions from {[}6{]}. The following table
summarizes the key abbreviations used throughout this paper.

\textbf{Table 1.} Notation: abbreviations and terminology.

\begin{longtable}[]{@{}lll@{}}
\toprule
Abbreviation & Expansion & Description\tabularnewline
\midrule
\endhead
NLZ & Number of Leading Zeros & Per-element hash rank\tabularnewline
\emph{B} & --- & Register (bucket) count\tabularnewline
\emph{V}, \emph{V\_t} & --- & Empty buckets total / at tier
\emph{t}\tabularnewline
CF & Correction Factor & Bias-correction lookup or
formula\tabularnewline
LC & Linear Counting & Classical empty-bucket estimator\tabularnewline
DLC & Dynamic Linear Counting & Tier-aware LC with info-power weighting
(§4.3)\tabularnewline
SBS & State Bias Sum & History-pattern LC for low cardinality
(§4.3)\tabularnewline
HC & History Counting & Per-tier estimator from history bits
(§4.3)\tabularnewline
DlcSbs & DLC--SBS blend & DLC + SBS blend for low-to-mid cardinality
(§4.2)\tabularnewline
LDLC & Layered DLC & DlcSbs + HC blend across cardinality
(§4.2)\tabularnewline
HLDLC & Hybrid LDLC & 50\% LDLC + 50\% Hybrid+2 --- primary estimator
(§4.1)\tabularnewline
MVP & Memory-Variance Product & RMSE² $\times$ memory in bits; lower is
better\tabularnewline
\bottomrule
\end{longtable}

\hypertarget{exaloglog}{%
\subsection{2.2 ExaLogLog}\label{exaloglog}}

ExaLogLog {[}1{]} extends UltraLogLog with variable-width registers of
6+\emph{t}+\emph{d} bits, where \emph{t} controls the NLZ precision
(sub-NLZ fractional bits) and \emph{d} controls the depth of sub-NLZ
history. The additional bits per register encode a more detailed picture
of the local hash distribution, which a maximum-likelihood (ML)
estimator exploits via Fisher-information analysis.

ExaLogLog achieves the best known memory-variance product among
fixed-register HLL variants. The memory-variance product (MVP) is
defined as RMSE² $\times$ (total memory in bits), where RMSE is the
root-mean-square relative error; a lower MVP indicates more efficient
use of memory. The theoretically optimal configuration is ELL(2,20) with
MVP = 3.67, but ELL(2,24) with MVP = 3.78 is preferred in practice
because its 32-bit registers align with machine word boundaries,
simplifying implementation and enabling efficient concurrent updates via
compare-and-swap {[}1{]}. We benchmark against ELL(2,24) as the
practical configuration. However, its registers are large: at these
parameters, each register occupies 32 bits. At 2 KB, this allows only
512 registers (\emph{p}=9), compared to HLL's 2,560 at 2 KB (packed 5
per 32-bit word). ExaLogLog compensates by extracting far more
information per register, but the register count disadvantage limits the
achievable accuracy per byte.

The ML estimator requires iterative numerical solving (Newton's method),
adding computational cost to each cardinality query. ExaLogLog also
supports a martingale estimator that avoids this cost but requires
observing every insertion, making it unsuitable for merge-and-estimate
workflows.

\hypertarget{ultraloglog-and-fgra}{%
\subsection{2.3 UltraLogLog and FGRA}\label{ultraloglog-and-fgra}}

UltraLogLog {[}5{]} uses 8-bit registers (6 NLZ + 2 history bits) with
the FGRA (Further Generalized Remaining Area) estimator. FGRA is a
closed-form estimator (no iterative solving) that achieves a 28\%
improvement in memory-variance product over HLL. At 2 KB (2,048
registers), ULL achieves 1.38\% width-weighted error --- a substantial
improvement over HLL's 1.65\%.

\hypertarget{dynamicloglog-and-estimation-hierarchy}{%
\subsection{2.4 DynamicLogLog and Estimation
Hierarchy}\label{dynamicloglog-and-estimation-hierarchy}}

DynamicLogLog and UltraDynamicLogLog {[}6{]} introduced several
innovations that AVLL builds upon:

\begin{itemize}
\tightlist
\item
  \textbf{Shared exponent} (\texttt{globalNLZ}): all registers store NLZ
  relative to a global floor, enabling 4-bit register storage and early
  exit. This technique was introduced earlier by Apache DataSketches
  HLL\_4 (\texttt{curMin}) and HyperLogLogLog {[}7{]} (shared base
  register); DLL's contribution is combining it with early exit and
  tier-aware estimation.
\item
  \textbf{Dynamic Linear Counting (DLC)}: tier-aware extension of LC
  that provides accurate estimates at any cardinality without a
  transition artifact.
\item
  \textbf{Estimation hierarchy}: a series of pairwise blends (DLC with
  HC, then with a corrected harmonic mean) that progressively reduce
  error through complementary bias cancellation. The companion tracker
  UDLL (6-bit registers with 2-bit history) further introduced State
  Bias Sum (SBS), which AVLL extends and fully describes here.
\end{itemize}

\hypertarget{avll-architecture}{%
\section{3. AVLL Architecture}\label{avll-architecture}}

\hypertarget{arithmetic-encoding}{%
\subsection{3.1 Arithmetic Encoding}\label{arithmetic-encoding}}

Traditional LogLog variants pack registers in fixed-width slots. HLL
uses 6-bit registers; in our packed implementation (LogLog30), 5
registers are stored per 32-bit word (30 of 32 bits used, 6.4 bits per
register including overhead) with modular bucket selection. DLL4 uses 4
bits (zero waste in 32-bit words), and ULL uses 8 bits (zero waste in
byte arrays). These fixed-width schemes require the per-register bit
width to evenly divide (or nearly divide) the word size for efficient
packing.

AVLL breaks this constraint. Each register can take one of 56 distinct
states (RADIX = 56), encoding both the NLZ exponent (0--18) and a 2-bit
sub-NLZ history pattern. Rather than storing each state in a fixed 6-bit
slot (which would waste capacity, since \(\lceil \log_2(56) \rceil = 6\)
but \(2^6 = 64 \neq 56\)), AVLL packs registers using mixed-radix
positional encoding:

\[\text{word} = \sum_{i=0}^{10} \text{reg}_i \times 56^i\]

Since
\(56^{11} \approx 1.70 \times 10^{19} < 2^{64} \approx 1.84 \times 10^{19}\),
eleven registers fit into a single unsigned 64-bit integer with no
padding bits --- all 64 bits participate in the encoding, utilizing
92.1\% of the 64-bit state space. This yields 64/11 = 5.82 bits per
register.

\textbf{Register access.} Reading register \emph{i}:

\begin{verbatim}
wordIdx = i / 11
pos = i % 11
reg = Long.remainderUnsigned(Long.divideUnsigned(word, POW[pos]), 56)
\end{verbatim}

Writing register \emph{i} (modular arithmetic update):

\begin{verbatim}
word += (newVal - oldVal) * POW[pos]
\end{verbatim}

where POW{[}k{]} = 56\^{}k is precomputed. The use of
\texttt{Long.divideUnsigned} and \texttt{Long.remainderUnsigned} ensures
correct handling of the full unsigned 64-bit range.

\textbf{Memory layout.} For \emph{B} registers, AVLL requires
\(\lceil B/11 \rceil\) 64-bit words. At 2,048 requested registers, this
rounds up to 187 words $\times$ 11 = 2,057 registers in 1,496 bytes. The actual
register count is always a multiple of 11; AVLL uses modular bucket
selection (rather than power-of-two masking) to distribute elements
across the non-power-of-two register count.

\textbf{Comparison with ExaLogLog.} At 2 KB, AVLL stores 2,816 registers
at 5.82 bits each, while ExaLogLog (at \emph{t}=2, \emph{d}=24,
\emph{p}=9) stores 512 registers at 32 bits each. AVLL has 5.5$\times$ more
registers. ExaLogLog's wider registers encode sub-NLZ fractional bits
(the \emph{t} lowest bits of the hash within the NLZ tier) and \emph{d}
bits of history, giving it richer per-register information. Which
approach extracts more total information from the same memory budget is
an empirical question --- and at every memory point tested, AVLL's
approach wins by 4--5\%.

\hypertarget{state-mapping}{%
\subsection{3.2 State Mapping}\label{state-mapping}}

\textbf{Table 2.} Register state encoding. Each register state (0--55)
encodes both an NLZ exponent and a 2-bit history pattern.

\begin{longtable}[]{@{}lllll@{}}
\toprule
Tier & States & Exponent & History bits & Notes\tabularnewline
\midrule
\endhead
T0 & 1 (state 0) & 0 & 0 & No observations above floor\tabularnewline
T1 & 2 (states 1--2) & 1 & MSB only & 1 valid history bit\tabularnewline
T2--T13 & 48 (states 3--50) & 2--13 & Both bits & Full 2-bit
history\tabularnewline
T14--T18 & 5 (states 51--55) & 14--18 & None & Above history tier
limit\tabularnewline
\bottomrule
\end{longtable}

The history tier limit (HTL = 13) determines the highest tier that
carries sub-NLZ history. Tiers 2 through 13 carry full 2-bit history (4
states each); tier 1 carries only 1 history bit (2 states); tier 0
carries none. States above HTL store only the exponent, trading history
for extended range. The maximum exponent (MAX\_EXP = 18) provides an
effective dynamic range of \(2^{18} \times B \approx 5.4 \times 10^{8}\)
at 2,048 registers within the current floor window --- the absolute
cardinality range depends on the hash width rather than the register
width, because the floor slides without bound (see Section 3.6).

The state-to-exponent and state-to-history mappings are computed
arithmetically (no lookup table required):

\[\text{exp}(s) = \begin{cases} 0 & s = 0 \\ 1 & s \in \{1, 2\} \\ \lfloor(s-3)/4\rfloor + 2 & 3 \leq s < 51 \\ 14 + (s - 51) & s \geq 51 \end{cases}\]

\hypertarget{hash-and-add}{%
\subsection{3.3 Hash and Add}\label{hash-and-add}}

AVLL uses Thomas Wang's 64-bit integer hash {[}8{]} and modular bucket
selection:

\begin{enumerate}
\def\labelenumi{\arabic{enumi}.}
\tightlist
\item
  Hash the input element with a shift-multiply finalizer.
\item
  Compare the hash code to a dynamic global minimum; return if it cannot
  alter any bucket.
\item
  Select the register index via
  \texttt{Long.remainderUnsigned(key,\ modBuckets)}.
\item
  Compute the NLZ and its position relative to the current global floor.
\item
  Unpack the existing register as a bitmap, set the new bit, and repack.
\item
  If the new state exceeds the old, update the register.
\end{enumerate}

The bitmap representation (one bit per NLZ position, with history bits
occupying positions NLZ-1 and NLZ-2) enables the register update to be
expressed as a bitwise OR followed by repacking --- the same approach
used by UltraLogLog for history bit updates.

\hypertarget{tier-promotion}{%
\subsection{3.4 Tier Promotion}\label{tier-promotion}}

Like DLL, AVLL maintains a global NLZ floor (\texttt{globalNLZ}) and a
floor counter tracking how many registers are at or near the floor. When
all registers have advanced beyond the floor zone (defined as exponent
\(\leq\) HISTORY\_MARGIN, a constant set to 2 that determines how many
sub-NLZ tiers are retained near the floor for history tracking), the
floor advances:

\begin{enumerate}
\def\labelenumi{\arabic{enumi}.}
\tightlist
\item
  Increment \texttt{globalNLZ}.
\item
  Demote all registers by one tier via \texttt{countAndDecrement()}.
\item
  Recount floor-zone registers.
\item
  Repeat if the floor is still saturated.
\end{enumerate}

Elements whose hash NLZ falls below the current floor cannot update any
register and are rejected before any register access. At high
cardinality, this rejects the vast majority of elements
(\textasciitilde96\% at 20M, \textgreater99.9\% beyond
\textasciitilde{}\(10^8\)), reducing AVLL's per-element cost to near
zero --- the same early exit mechanism as DLL4 {[}6{]}. This mechanism
is responsible for AVLL's throughput advantage under cache pressure
(§6.5).

\hypertarget{merge}{%
\subsection{3.5 Merge}\label{merge}}

AVLL supports register-wise merge for combining sketches from parallel
threads. Each register is unpacked into its bitmap representation, the
bitmaps are OR'd (with appropriate shifting to align different global
NLZ floors), and the result is repacked. This preserves both the NLZ
exponent and history bits, though the merged result may have slightly
higher overflow than a single-stream instance due to independent tier
promotion (see {[}6{]} for quantification).

\hypertarget{cardinality-range-and-memory-scaling}{%
\subsection{3.6 Cardinality Range and Memory
Scaling}\label{cardinality-range-and-memory-scaling}}

The shared-exponent architecture that makes AVLL non-idempotent also
gives it two properties that fixed-register schemes lack.

\textbf{No cardinality ceiling.} In HLL, ExaLogLog, and other
fixed-register schemes, each register stores an absolute NLZ value. The
register width bounds the maximum representable NLZ, which in turn
bounds the maximum representable cardinality. With 6-bit HLL registers
and 64-bit hashes, the ceiling is approximately
\(2^{64} \approx 1.8 \times 10^{19}\). ExaLogLog with 32-bit registers
faces the same hash-space ceiling. Neither architecture can benefit from
wider hash functions: even with a 256-bit hash, a 6-bit register can
still only store NLZ values 0--63.

AVLL stores \emph{relative} NLZ values (absolute NLZ minus the global
floor \texttt{globalNLZ}), and the per-register range is fixed at 0--18
regardless of cardinality. The \texttt{globalNLZ} counter is a 32-bit
integer that slides upward as cardinality increases. The maximum
representable cardinality is therefore bounded by the hash width, not
the register width: with a 256-bit hash, AVLL can represent
cardinalities up to approximately \(2^{256} \approx 10^{77}\). Since
\texttt{globalNLZ} can trivially be widened to 64 bits if needed, and
wider hashes can be constructed by concatenating multiple independent
hash functions, AVLL imposes no practical cardinality ceiling.

\textbf{Cardinality-independent memory.} Fixed-register schemes require
O(\emph{B} $\times$ log log \emph{C}) memory because each of \emph{B} registers
must be wide enough to store NLZ values up to
\(\log_2(C/B) \approx \log_2 C\), requiring \(\log_2 \log_2 C\) bits per
register. AVLL's per-register width is constant (5.82 bits, encoding 56
states independent of cardinality), with only the single global
\texttt{globalNLZ} counter growing with \emph{C}. The total memory is
therefore O(\emph{B} + log log \emph{C}) --- the cardinality term is
additive rather than multiplicative, making \emph{C} effectively
irrelevant to the memory budget. At any practical cardinality, the
\texttt{globalNLZ} counter contributes 4 bytes to a sketch that is
otherwise hundreds or thousands of bytes, so the memory cost is
dominated entirely by the register count \emph{B}.

\hypertarget{hldlc-estimation-pipeline}{%
\section{4. HLDLC Estimation Pipeline}\label{hldlc-estimation-pipeline}}

\hypertarget{the-hldlc-estimator}{%
\subsection{4.1 The HLDLC Estimator}\label{the-hldlc-estimator}}

AVLL's primary estimator, HLDLC (Hybrid Layered DLC), blends two
fundamentally different estimation frameworks with equal weights:

\[\text{HLDLC} = 0.50 \times \text{LDLC} + 0.50 \times \text{Hybrid{+}2}\]

LDLC (Layered DLC) is an LC-based framework: it combines tier-aware
linear counting (DLC) with state bias sum (SBS) at low cardinality and
history counting (HC) at high cardinality. Hybrid+2 is a
harmonic-mean-based framework: it combines a history-corrected harmonic
mean (Mean+H) at high cardinality with SBS at low cardinality. The
residual errors of these two frameworks are largely uncorrelated in the
mid-to-high cardinality regime where the component estimators diverge.
At very low cardinality, both branches reduce to SBS-only estimation and
their errors are fully correlated, but this has no practical impact
because SBS is already the optimal estimator (of the components) in that
range. The 50/50 blend achieves complementary bias cancellation that
reduces error below any single-framework estimator.

The estimation hierarchy is:

\begin{verbatim}
HLDLC = 50% LDLC + 50% Hybrid+2
  where LDLC    = blend(DlcSbs, HC)      --- LC-based + history-based
        DlcSbs  = blend(DLC, SBS)        --- tier-aware LC + state-aware LC
        Hybrid+2 = blend(SBS, Mean+H)    --- state-aware LC -> harmonic mean
\end{verbatim}

\textbf{Table 3.} Accuracy of each HLDLC component at 1 KB (1,408
registers), high-complexity data, 128,000 instances. WidthWtAbsErr is
the primary metric.

\begin{longtable}[]{@{}lllll@{}}
\toprule
Component & LogWtAbsErr\% & WidthWtAbsErr\% & PeakAbsErr\% & Accurate
Range\tabularnewline
\midrule
\endhead
Mean+H & 2.08 & 1.67 & 114 & Mid-high card\tabularnewline
SBS & 44.97 & 99.14 & 99.89 & Low card only\tabularnewline
DLC & 1.93 & 2.25 & 2.26 & Full range\tabularnewline
HC & 20.22 & 2.39 & 181 & High card only\tabularnewline
DlcSbs & --- & --- & --- & Full range\tabularnewline
LDLC & 1.39 & 1.64 & 1.65 & Full range\tabularnewline
Hybrid+2 & 1.42 & 1.67 & 1.67 & Full range\tabularnewline
\textbf{HLDLC} & \textbf{1.38} & \textbf{1.63} & \textbf{1.64} &
\textbf{Full range}\tabularnewline
\bottomrule
\end{longtable}

Mean+H, SBS, and HC are accurate only in their stated cardinality range
and serve as inputs to blends, not as standalone estimators. Peak values
exceeding 100\% (Mean+H: 114\%, HC: 181\%) indicate the component
diverges outside its useful range --- this is expected for narrow-range
components. DlcSbs is an internal intermediate (DLC blended with SBS)
that is not instrumented separately in the calibration pipeline; it is
listed for hierarchy completeness.

All blend thresholds in the HLDLC hierarchy were determined by
minimizing width-weighted error over 512,000 independent simulations of
8,192 $\times$ \emph{B} unique elements each (approximately 16.8 million at
\emph{B} = 2,048). The thresholds are insensitive to small
perturbations: varying any threshold by ±20\% affects only log-weighted
error (which emphasizes low-cardinality behavior) and has negligible
effect on width-weighted error.

\hypertarget{the-two-branches}{%
\subsection{4.2 The Two Branches}\label{the-two-branches}}

\textbf{LDLC (Layered DLC)} blends the DLC--SBS component (DlcSbs) with
HC across the cardinality range using a linear ramp:

\[\text{LDLC} = \begin{cases} \hat{n}_{\text{DlcSbs}} & \hat{n}_{\text{DlcSbs}} \leq 0.5B \\ (1 - w_{\text{HC}})\,\hat{n}_{\text{DlcSbs}} + w_{\text{HC}}\,\hat{n}_{\text{HC}} & 0.5B < \hat{n}_{\text{DlcSbs}} \leq 4.5B \\ 0.50\,\hat{n}_{\text{DlcSbs}} + 0.50\,\hat{n}_{\text{HC}} & \hat{n}_{\text{DlcSbs}} > 4.5B \end{cases}\]

where
\(w_{\text{HC}} = \frac{\hat{n}_{\text{DlcSbs}} - 0.5B}{4B} \times 0.50\)
ramps the HC weight from 0 to a maximum of 50\%. If HC produces no valid
tier estimates (fewer than 8 effective buckets at all tiers), LDLC falls
back to the DLC--SBS blend alone. The complementary bias structure of
these two components produces cancellation that reduces periodic error
by 27\% width-weighted relative to DLC alone {[}6{]}.

\textbf{DlcSbs} blends DLC with SBS using logarithmic interpolation:

\[\text{DlcSbs} = \begin{cases} \hat{n}_{\text{SBS}} & \hat{n}_{\text{DLC}} \leq 2B \\ (1-\tau)\,\hat{n}_{\text{SBS}} + \tau\,\hat{n}_{\text{DLC}} & 2B < \hat{n}_{\text{DLC}} < 6B \\ \hat{n}_{\text{DLC}} & \hat{n}_{\text{DLC}} \geq 6B \end{cases}\]

where \(\tau = \ln(\hat{n}_{\text{DLC}} / (2B)) / \ln(3)\). This
captures SBS's superior accuracy at very low cardinalities and DLC's
robust accuracy at higher cardinalities.

\textbf{Hybrid+2} blends SBS at low cardinality with Mean+H at high
cardinality, using DLC's raw estimate as a cardinality reference:

\[\text{Hybrid{+}2} = \begin{cases} \hat{n}_{\text{SBS}} & \hat{n}_{\text{DLC}} \leq 1B \\ (1-\tau)\,\hat{n}_{\text{SBS}} + \tau\,\hat{n}_{\text{MeanH}} & 1B < \hat{n}_{\text{DLC}} < 6B \\ \hat{n}_{\text{MeanH}} & \hat{n}_{\text{DLC}} \geq 6B \end{cases}\]

where \(\tau = \ln(\hat{n}_{\text{DLC}} / B) / \ln(6)\). The ``+2''
indicates that it uses Mean+H (the 2-bit history-enhanced Mean) rather
than plain Mean.

Throughout the hierarchy, DLC's raw output serves as the cardinality
reference for all blend-zone detection and CF seed values.

\hypertarget{the-four-components}{%
\subsection{4.3 The Four Components}\label{the-four-components}}

Each component exploits a different aspect of the register state. We
describe them in order of their primary cardinality regime, matching
Figure 4's visual story.

\textbf{State Bias Sum (SBS)} --- \emph{low cardinality; consumed by
DlcSbs and Hybrid+2; first complete description here (SBS was
implemented in UDLL {[}6{]} but not described in detail; the DLL paper
mentions it by name only).}

SBS exploits the full register state --- not just empty vs.~filled, but
the specific 2-bit history pattern --- to estimate cardinality where
Linear Counting has high variance. Each register is mapped to one of 11
state indices based on its NLZ bin and history pattern (Table 4).

\textbf{Table 4.} SBS state-index mapping.

\begin{longtable}[]{@{}llll@{}}
\toprule
NLZ bin & Valid history bits & States & Indices\tabularnewline
\midrule
\endhead
0 & 0 & 1 (empty) & 0\tabularnewline
1 & 1 (MSB only) & 2 & 1--2\tabularnewline
2 & 2 (both bits) & 4 & 3--6\tabularnewline
3+ & 2 (both bits) & 4 & 7--10\tabularnewline
\bottomrule
\end{longtable}

For each contributing register, a cubic polynomial in the occupancy
parameter \(L = \ln(B / \max(0.5, B - F_s))\) computes the expected
distinct count, where \(F_s\) is the count of non-empty registers with
structurally valid SBS states (registers with impossible history
patterns are excluded, so \(F_s \leq F\) where \(F\) is the total filled
register count):

\[\text{SBS}(s_i) = \text{base}[s_i] + a[s_i] \cdot L + b[s_i] \cdot L^2 + c[s_i] \cdot L^3\]

The total SBS estimate is
\(\hat{n}_{\text{SBS}} = \max\!\left(\sum_i \text{SBS}(s_i),\, F\right)\),
floored at the filled register count \(F\). SBS is accurate only below
approximately 6$\times$\emph{B}; above this, the polynomial model diverges. The
44 polynomial coefficients are bucket-count-independent
(\(R^2 \geq 0.999\); see §4.4 for details).

\textbf{Dynamic Linear Counting (DLC)} --- \emph{full range backbone;
consumed by DlcSbs and as cardinality reference; inherited from
{[}6{]}.}

DLC extends classical Linear Counting by computing tier-aware empty
counts. For each tier \emph{t}, a separate LC estimate is computed:

\[\text{DLC}(t) = 2^t \cdot B \cdot \ln\!\left(\frac{B}{\max(V_t, 0.5)}\right)\]

where \(V_t\) is the cumulative count of registers with NLZ exponent
\(\leq\) \emph{t}. Participating tiers (those with \(V_t\) in a reliable
range) are blended using inverse-error weighted log-space averaging with
sharpening exponent 4.5. DLC achieves 2.25\% width-weighted error
standalone and eliminates the LC-to-LogLog transition artifact that
plagues standard HLL. DLC is accurate across all cardinalities with no
correction factor, and thus is useful for calculating correction factors
or blend ratios for other components.

\textbf{History Counting (HC)} --- \emph{high cardinality; bias
complementary to DlcSbs; consumed by LDLC; inherited from {[}6{]}.}

HC uses the history bits of registers to construct independent
cardinality estimates at each tier boundary. For each tier \emph{t}, HC
examines registers at tiers \emph{t}+1 through \emph{t}+\emph{H} (where
\emph{H} = 2) and counts how many have each history bit set. The
per-tier estimate is:

\[\hat{n}_{\text{HC}}(t) = 2^{t+1} \cdot B \cdot \ln\!\left(\frac{B_{\text{eff}}(t)}{U(t)}\right)\]

where \(B_{\text{eff}}(t) = \sum_{d=0}^{H-1} C_{t+d+1}\) is the
effective bucket count (the number of registers at tiers \(t\)+1 through
\(t\)+\(H\)), \(U(t) = \sum_{d=0}^{H-1} (C_{t+d+1} - S_{d,t+d+1})\) is
the unseen count (\(C_k\) = registers at tier \(k\), \(S_{d,k}\) =
registers at tier \(k\) with history bit \(d\) set), and a per-tier
estimate requires \(B_{\text{eff}} \geq 8\) and
\(1 \leq U < B_{\text{eff}}\). Per-tier estimates are blended using
inverse-error weighted log-space averaging (see {[}6{]} for full
derivation). HC's critical property is that its bias structure is
complementary to DlcSbs: where DlcSbs overestimates, HC tends to
underestimate, enabling the cancellation that LDLC exploits.

\textbf{Mean (foundation for Mean+H, not a standalone HLDLC component)}
--- \emph{classical lineage with occupancy correction.}

The Mean estimator computes an occupancy-corrected harmonic mean of
register values. For each register with absolute NLZ exponent \emph{e},
the contribution is \(\text{dif}(e) = 2^{63-e}\). The raw estimate is:

\[\hat{n}_{\text{Mean}} = 2 \cdot \frac{M_{\max}}{\bar{d}} \cdot F \cdot c \cdot T_{\text{Mean}}\]

where \(M_{\max} = 2^{63}-1\),
\(\bar{d} = \frac{1}{F}\sum_i \text{dif}(e_i)\) is the average
contribution over the \(F\) filled registers, \(c = (F+B)/(2B)\) is the
occupancy correction, and \(T_{\text{Mean}} = 0.721\) is the terminal
correction factor. Mean is most accurate at high cardinality (2.22\%
width-weighted) and serves as the foundation for Mean+H, not as a
standalone estimator.

\textbf{Mean+History (Mean+H)} --- \emph{mid-to-high cardinality;
consumed by Hybrid+2; novel history-offset mechanism.}

Mean+H extends Mean by incorporating the 2-bit sub-NLZ history pattern
of each register. Each register's contribution is adjusted by a
per-tier, per-history-pattern offset:

\[\text{corrDif}(e_i, b_i, h_i) = \text{dif}(e_i) \cdot 2^{-(\delta(b_i, h_i) + \delta_0)} \cdot T_{\text{MeanH}}^{-1}\]

where \(\delta(b, h)\) is the history offset lookup and
\(T_{\text{MeanH}} = 0.856\) is the terminal correction factor. The
offsets encode how much additional information the history pattern
provides about the register's true cardinality contribution. Mean+H
reduces width-weighted error from 2.22\% (Mean) to 1.67\% --- a 25\%
improvement from history-bit exploitation.

\hypertarget{correction-formulas}{%
\subsection{4.4 Correction Formulas}\label{correction-formulas}}

AVLL embeds all correction formulas as static constants in the Java
class. This section details the formulas for implementers; the preceding
sections provide all information needed to understand the estimation
pipeline conceptually.

\textbf{Mean and Mean+H correction factors.} Both use a 3-Sigmoid +
2-Gaussian (3S2G) formula with 16 coefficients (R² = 0.9999):

\[\text{CF}(x) = a_0 + a_1 S(x, c_1, w_1) + a_2 S(x, c_2, w_2) + a_3 S(x, c_3, w_3) + g_1 G(x, \mu_1, \sigma_1) + g_2 G(x, \mu_2, \sigma_2)\]

where \(x = \log_2(\hat{n}/B)\),
\(S(x,c,w) = \frac{1}{2}(1+\tanh(\frac{x-c}{w}))\), and
\(G(x,\mu,\sigma) = \exp(-(\frac{x-\mu}{\sigma})^2)\). The CF is
evaluated using DLC's raw estimate as the seed; when the seed exceeds
10$\times$\emph{B}, a single refinement step re-evaluates using the corrected
estimate (at most two evaluations total). Mean and Mean+H use different
coefficient sets but the same functional form.

\textbf{HC correction factor.} An exponential + sinusoidal formula (7
coefficients) that models the periodic bias structure inherent in
tier-boundary estimation:

\[\text{CF}_{\text{HC}}(x) = T - A \cdot e^{-Bx} + (S_0 + S_1 x)\sin(2\pi x) + (C_0 + C_1 x)\cos(2\pi x)\]

where \(x = \log_2(\hat{n}_{\text{HC}})\).

\textbf{SBS polynomial coefficients.} 44 coefficients (4 per state $\times$ 11
states), fitted across bucket counts 256--2,048. The coefficients are
bucket-count-independent (\(R^2 \geq 0.999\)), so a single set serves
all configurations. Registers with structurally impossible history
patterns (nonzero bits in positions that should be zero) are excluded.

\textbf{History offset table.} 16 values (3 per-tier arrays $\times$ 4 history
states for tiers 0--2, plus 4 steady-state values for tiers 3+) that
encode per-register corrections for Mean+H.

\textbf{Table 5.} Embedded constants summary.

\begin{longtable}[]{@{}lll@{}}
\toprule
\begin{minipage}[b]{0.29\columnwidth}\raggedright
Formula\strut
\end{minipage} & \begin{minipage}[b]{0.36\columnwidth}\raggedright
Parameters\strut
\end{minipage} & \begin{minipage}[b]{0.26\columnwidth}\raggedright
Source\strut
\end{minipage}\tabularnewline
\midrule
\endhead
\begin{minipage}[t]{0.29\columnwidth}\raggedright
SBS polynomial\strut
\end{minipage} & \begin{minipage}[t]{0.36\columnwidth}\raggedright
44 coefficients (11 states $\times$ 4 terms)\strut
\end{minipage} & \begin{minipage}[t]{0.26\columnwidth}\raggedright
Cubic fit, B-independent\strut
\end{minipage}\tabularnewline
\begin{minipage}[t]{0.29\columnwidth}\raggedright
Mean CF (3S2G)\strut
\end{minipage} & \begin{minipage}[t]{0.36\columnwidth}\raggedright
16 coefficients\strut
\end{minipage} & \begin{minipage}[t]{0.26\columnwidth}\raggedright
Sigmoid+Gaussian fit on simulation\strut
\end{minipage}\tabularnewline
\begin{minipage}[t]{0.29\columnwidth}\raggedright
Mean+H CF (3S2G)\strut
\end{minipage} & \begin{minipage}[t]{0.36\columnwidth}\raggedright
16 coefficients\strut
\end{minipage} & \begin{minipage}[t]{0.26\columnwidth}\raggedright
Same framework, different coefficients\strut
\end{minipage}\tabularnewline
\begin{minipage}[t]{0.29\columnwidth}\raggedright
HC CF (exp+sin)\strut
\end{minipage} & \begin{minipage}[t]{0.36\columnwidth}\raggedright
7 coefficients\strut
\end{minipage} & \begin{minipage}[t]{0.26\columnwidth}\raggedright
Exponential + sinusoidal fit\strut
\end{minipage}\tabularnewline
\begin{minipage}[t]{0.29\columnwidth}\raggedright
History offsets\strut
\end{minipage} & \begin{minipage}[t]{0.36\columnwidth}\raggedright
16 values (3 per-tier + 1 steady-state, each $\times$ 4 states)\strut
\end{minipage} & \begin{minipage}[t]{0.26\columnwidth}\raggedright
Per-tier correction for Mean+H\strut
\end{minipage}\tabularnewline
\bottomrule
\end{longtable}

Total: approximately 100 embedded constants. No external files, no
runtime dependencies. This makes AVLL fully self-contained --- a single
Java class plus the CardinalityTracker interface.

\hypertarget{methods}{%
\section{5. Methods}\label{methods}}

\hypertarget{simulation-framework}{%
\subsection{5.1 Simulation Framework}\label{simulation-framework}}

All accuracy evaluations used the same simulation infrastructure as
{[}6{]}: purpose-built calibration drivers in BBTools running 128,000
independent estimator instances, sampling at exponentially-spaced
cardinality checkpoints (1\% multiplicative increments, i.e., each
checkpoint is 1.01$\times$ the previous), with results merged across all
instances. Calibration (correction factor fitting) and evaluation used
independent random seeds --- each estimator instance receives a unique
seed derived from a master seed, and the evaluation master seed differs
from the calibration master seed. The LC distribution shift (a
100-million-element pool with skewed duplicates, yielding
\textasciitilde87 million observed distinct values after 400 million
additions, vs.~all-unique HC streams) provides additional out-of-sample
evidence that the correction formulas generalize beyond the calibration
distribution. ExaLogLog was tested via a clean-room reimplementation of
the algorithms described in {[}1{]}, as the reference implementation's
license prohibits redistribution. Our reimplementation uses the same
default parameters (\emph{t}=2, \emph{d}=24, 32-bit registers) and ML
estimation algorithm (Newton's method on the log-likelihood function) as
described in {[}1{]}.

\textbf{Table 6.} ExaLogLog validation. We validated our
reimplementation against Ertl's theoretical predictions. For ELL(2,24),
the theoretical memory-variance product is MVP = 3.78, giving a
predicted RMSE of \(\sqrt{\text{MVP}/((q+d)m)}\) where q = 6+t = 8 and m
is the number of registers. Since our primary metric is width-weighted
mean absolute error (MAE) rather than RMSE, we compare against the
theoretical expected MAE = \(\sqrt{2/\pi} \times\) RMSE
\(\approx 0.798 \times\) RMSE, which is exact for Gaussian-distributed
relative errors.

\begin{longtable}[]{@{}llllll@{}}
\toprule
Memory & Registers (\emph{m}) & Ertl RMSE (theory) & Expected MAE & Our
WidthWtAbsErr & Ratio\tabularnewline
\midrule
\endhead
0.25 KB & 64 & 4.30\% & 3.43\% & 3.42\% & 0.997\tabularnewline
0.5 KB & 128 & 3.04\% & 2.42\% & 2.42\% & 1.000\tabularnewline
1 KB & 256 & 2.15\% & 1.71\% & 1.71\% & 0.999\tabularnewline
2 KB & 512 & 1.52\% & 1.21\% & 1.21\% & 0.999\tabularnewline
4 KB & 1024 & 1.07\% & 0.86\% & 0.86\% & 0.998\tabularnewline
\bottomrule
\end{longtable}

Our measured error matches the theoretical prediction within 0.3\%
relative across all five memory points, confirming that the
reimplementation faithfully reproduces Ertl's algorithms. At 1 KB (p=8,
256 registers), Ertl's paper reports an empirical RMSE of 2.15\% from 1
million simulation runs {[}1{]}, consistent with the theoretical value
of 2.148\% --- and our WidthWtAbsErr of 1.71\% matches the corresponding
expected MAE of 1.71\%.

\textbf{High-complexity (HC) mode.} All-unique elements from
Xoshiro256++ PRNG, hashed through Thomas Wang's 64-bit finalizer. Each
estimator instance processes all-unique elements until the true
cardinality reaches 8,192 $\times$ \emph{B} (the bucket count), then is
discarded.

\textbf{Low-complexity (LC) mode.} Elements drawn with replacement from
a bounded pool of 100 million unique values via array{[}min(rand(),
rand())*arrayLen{]} to simulate skewed frequency distributions. Each
estimator instance replays the value pool 4 times (iter=4), drawing with
replacement within each replay, for a total of exactly 400 million
additions per instance at \textasciitilde87 million distinct elements;
32,000 independent instances. Tests robustness under sustained
duplication.

Both modes report three error metrics: - \textbf{Width-weighted mean
absolute error (WidthWtAbsErr)}: the primary metric. Error at each
checkpoint is weighted by the cardinality interval width, reflecting a
linearly-distributed workload. - \textbf{Log-weighted mean absolute
error (LogWtAbsErr)}: each log-spaced checkpoint weighted equally.
Emphasizes low-cardinality behavior. - \textbf{Peak absolute error
(PeakAbsErr)}: worst-case error at any single checkpoint.

\hypertarget{memory-equivalent-comparison-methodology}{%
\subsection{5.2 Memory-Equivalent Comparison
Methodology}\label{memory-equivalent-comparison-methodology}}

Comparing cardinality estimators at equal memory requires careful
accounting of per-register bit widths. Each estimator type was
configured with the maximum register count that fits within a given
memory budget (Table 7).

\textbf{Table 7.} Estimator configurations at 1 KB memory budget.

\begin{longtable}[]{@{}lllll@{}}
\toprule
\begin{minipage}[b]{0.15\columnwidth}\raggedright
Type\strut
\end{minipage} & \begin{minipage}[b]{0.09\columnwidth}\raggedright
Bits/reg\strut
\end{minipage} & \begin{minipage}[b]{0.33\columnwidth}\raggedright
Encoding\strut
\end{minipage} & \begin{minipage}[b]{0.11\columnwidth}\raggedright
1 KB registers\strut
\end{minipage} & \begin{minipage}[b]{0.18\columnwidth}\raggedright
Storage\strut
\end{minipage}\tabularnewline
\midrule
\endhead
\begin{minipage}[t]{0.15\columnwidth}\raggedright
AVLL\strut
\end{minipage} & \begin{minipage}[t]{0.09\columnwidth}\raggedright
5.82\strut
\end{minipage} & \begin{minipage}[t]{0.33\columnwidth}\raggedright
Base-56 arithmetic, 11 per 64-bit word\strut
\end{minipage} & \begin{minipage}[t]{0.11\columnwidth}\raggedright
1,408\strut
\end{minipage} & \begin{minipage}[t]{0.18\columnwidth}\raggedright
\texttt{long{[}{]}}\strut
\end{minipage}\tabularnewline
\begin{minipage}[t]{0.15\columnwidth}\raggedright
ExaLogLog\strut
\end{minipage} & \begin{minipage}[t]{0.09\columnwidth}\raggedright
32\strut
\end{minipage} & \begin{minipage}[t]{0.33\columnwidth}\raggedright
\emph{t}=2, \emph{d}=24, one register per \texttt{int}\strut
\end{minipage} & \begin{minipage}[t]{0.11\columnwidth}\raggedright
256\strut
\end{minipage} & \begin{minipage}[t]{0.18\columnwidth}\raggedright
\texttt{int{[}{]}}\strut
\end{minipage}\tabularnewline
\begin{minipage}[t]{0.15\columnwidth}\raggedright
HLL-TailCut4\strut
\end{minipage} & \begin{minipage}[t]{0.09\columnwidth}\raggedright
4\strut
\end{minipage} & \begin{minipage}[t]{0.33\columnwidth}\raggedright
4-bit offset + shared base register\strut
\end{minipage} & \begin{minipage}[t]{0.11\columnwidth}\raggedright
2,048\strut
\end{minipage} & \begin{minipage}[t]{0.18\columnwidth}\raggedright
\texttt{byte{[}{]}}\strut
\end{minipage}\tabularnewline
\begin{minipage}[t]{0.15\columnwidth}\raggedright
HyperLogLogLog\strut
\end{minipage} & \begin{minipage}[t]{0.09\columnwidth}\raggedright
3+vc\strut
\end{minipage} & \begin{minipage}[t]{0.33\columnwidth}\raggedright
3-bit offset + exception map (unbounded)\strut
\end{minipage} & \begin{minipage}[t]{0.11\columnwidth}\raggedright
2,048\strut
\end{minipage} & \begin{minipage}[t]{0.18\columnwidth}\raggedright
\texttt{byte{[}{]}} + \texttt{IntHashMap}\strut
\end{minipage}\tabularnewline
\begin{minipage}[t]{0.15\columnwidth}\raggedright
Apache HLL4\strut
\end{minipage} & \begin{minipage}[t]{0.09\columnwidth}\raggedright
4+vc\strut
\end{minipage} & \begin{minipage}[t]{0.33\columnwidth}\raggedright
4-bit nibbles packed 2/byte + overflow map (unbounded)\strut
\end{minipage} & \begin{minipage}[t]{0.11\columnwidth}\raggedright
2,048\strut
\end{minipage} & \begin{minipage}[t]{0.18\columnwidth}\raggedright
\texttt{byte{[}{]}} + \texttt{HashMap}\strut
\end{minipage}\tabularnewline
\begin{minipage}[t]{0.15\columnwidth}\raggedright
HyperTwoBits\strut
\end{minipage} & \begin{minipage}[t]{0.09\columnwidth}\raggedright
2\strut
\end{minipage} & \begin{minipage}[t]{0.33\columnwidth}\raggedright
2-bit packed in \texttt{long{[}{]}} (MSB+LSB arrays)\strut
\end{minipage} & \begin{minipage}[t]{0.11\columnwidth}\raggedright
4,096\strut
\end{minipage} & \begin{minipage}[t]{0.18\columnwidth}\raggedright
\texttt{long{[}{]}} $\times$ 2\strut
\end{minipage}\tabularnewline
\begin{minipage}[t]{0.15\columnwidth}\raggedright
UltraLogLog\strut
\end{minipage} & \begin{minipage}[t]{0.09\columnwidth}\raggedright
8\strut
\end{minipage} & \begin{minipage}[t]{0.33\columnwidth}\raggedright
6 NLZ + 2 history, one per byte\strut
\end{minipage} & \begin{minipage}[t]{0.11\columnwidth}\raggedright
1,024\strut
\end{minipage} & \begin{minipage}[t]{0.18\columnwidth}\raggedright
\texttt{byte{[}{]}}\strut
\end{minipage}\tabularnewline
\begin{minipage}[t]{0.15\columnwidth}\raggedright
LogLog30\strut
\end{minipage} & \begin{minipage}[t]{0.09\columnwidth}\raggedright
6.4\strut
\end{minipage} & \begin{minipage}[t]{0.33\columnwidth}\raggedright
6-bit packed 5 per 32-bit word\strut
\end{minipage} & \begin{minipage}[t]{0.11\columnwidth}\raggedright
1,280\strut
\end{minipage} & \begin{minipage}[t]{0.18\columnwidth}\raggedright
\texttt{int{[}{]}}\strut
\end{minipage}\tabularnewline
\bottomrule
\end{longtable}

Register counts for HLL-family estimators (HTC4, HLLL, HLL4) are
constrained to powers of two by their hash-masking bucket selection,
which may leave unused memory within the budget. Each type uses its best
available estimator: HLDLC for AVLL, ML for ExaLogLog, and the default
(typically harmonic mean with LC blend) for all others.

\hypertarget{throughput-benchmark-protocol}{%
\subsection{5.3 Throughput Benchmark
Protocol}\label{throughput-benchmark-protocol}}

Insertion throughput was measured using two dedicated benchmark
harnesses (SpeedTest and SpeedTest2) that time only \texttt{add()}
operations --- no estimation checkpoints, no statistics accumulation.
This isolates the per-element insertion cost from the estimation
overhead that dominates in calibration runs.

\textbf{SpeedTest} (ALU-constrained): each thread creates one estimator
at a time, adds \texttt{card} elements, calls \texttt{cardinality()},
and discards. The estimator's working set fits in L1/L2 cache
throughout. Measures peak per-element throughput.

\textbf{SpeedTest2} (bandwidth-constrained): each thread maintains
\texttt{sim} simultaneous estimators in an array and cycles through all
of them on every add. This creates cache pressure proportional to
\texttt{sim} $\times$ sketch size $\times$ threads. At sim=4,096 with 128 threads and
2 KB per sketch, the total working set exceeds 1 GB --- far larger than
L3 cache on the processor.

All throughput benchmarks were run on a single 64-core cluster node (AMD
EPYC 9555P, 128 hardware threads, 256 MB L3 cache) at 2 KB equivalent
memory, matching the equal-memory methodology of §5.2. Each type was
tested independently to avoid cross-type interference. The \texttt{card}
parameter was set high enough that each type ran for at least 10 seconds
total, ensuring JVM warmup artifacts are negligible: 40M elements per
estimator for ALU mode, 10M for sim=2,048, and 4M for sim=4,096.
Run-to-run throughput variance across independent executions on the same
node type is approximately 10\%; throughput comparisons below this
margin should be interpreted as ties.

\textbf{Note on victim caches.} HyperLogLogLog and Apache HLL4 both use
unbounded auxiliary maps (IntHashMap and HashMap respectively) to store
register values that exceed their small fixed-width registers. At 1 KB
nominal memory, these maps may use additional memory beyond the stated
budget. This is inherent to these algorithms' designs. AVLL, ExaLogLog,
HLL-TailCut4, and HyperTwoBits use only their fixed-size register arrays
with no auxiliary storage.

\hypertarget{experimental-results}{%
\section{6. Experimental Results}\label{experimental-results}}

\hypertarget{avll-vs-exaloglog-memory-scaling-comparison-high-complexity}{%
\subsection{6.1 AVLL vs ExaLogLog: Memory-Scaling Comparison (High
Complexity)}\label{avll-vs-exaloglog-memory-scaling-comparison-high-complexity}}

\textbf{Table 8.} AVLL HLDLC vs ExaLogLog ML under high-complexity
conditions (all-unique elements), 128,000 independent instances. LogWt =
log-weighted mean absolute error; WidthWt = width-weighted mean absolute
error (primary metric); Peak = worst-case error. All values in percent.

\begin{longtable}[]{@{}llllllll@{}}
\toprule
\begin{minipage}[b]{0.07\columnwidth}\raggedright
Memory\strut
\end{minipage} & \begin{minipage}[b]{0.12\columnwidth}\raggedright
AVLL WidthWt\strut
\end{minipage} & \begin{minipage}[b]{0.12\columnwidth}\raggedright
EXA WidthWt\strut
\end{minipage} & \begin{minipage}[b]{0.10\columnwidth}\raggedright
AVLL LogWt\strut
\end{minipage} & \begin{minipage}[b]{0.10\columnwidth}\raggedright
EXA LogWt\strut
\end{minipage} & \begin{minipage}[b]{0.09\columnwidth}\raggedright
AVLL Peak\strut
\end{minipage} & \begin{minipage}[b]{0.08\columnwidth}\raggedright
EXA Peak\strut
\end{minipage} & \begin{minipage}[b]{0.10\columnwidth}\raggedright
WidthWt \(\Delta\)\strut
\end{minipage}\tabularnewline
\midrule
\endhead
\begin{minipage}[t]{0.07\columnwidth}\raggedright
0.25 KB\strut
\end{minipage} & \begin{minipage}[t]{0.12\columnwidth}\raggedright
3.28\strut
\end{minipage} & \begin{minipage}[t]{0.12\columnwidth}\raggedright
3.42\strut
\end{minipage} & \begin{minipage}[t]{0.10\columnwidth}\raggedright
2.86\strut
\end{minipage} & \begin{minipage}[t]{0.10\columnwidth}\raggedright
2.90\strut
\end{minipage} & \begin{minipage}[t]{0.09\columnwidth}\raggedright
3.29\strut
\end{minipage} & \begin{minipage}[t]{0.08\columnwidth}\raggedright
3.43\strut
\end{minipage} & \begin{minipage}[t]{0.10\columnwidth}\raggedright
-4.1\%\strut
\end{minipage}\tabularnewline
\begin{minipage}[t]{0.07\columnwidth}\raggedright
0.5 KB\strut
\end{minipage} & \begin{minipage}[t]{0.12\columnwidth}\raggedright
2.31\strut
\end{minipage} & \begin{minipage}[t]{0.12\columnwidth}\raggedright
2.42\strut
\end{minipage} & \begin{minipage}[t]{0.10\columnwidth}\raggedright
1.99\strut
\end{minipage} & \begin{minipage}[t]{0.10\columnwidth}\raggedright
2.02\strut
\end{minipage} & \begin{minipage}[t]{0.09\columnwidth}\raggedright
2.32\strut
\end{minipage} & \begin{minipage}[t]{0.08\columnwidth}\raggedright
2.44\strut
\end{minipage} & \begin{minipage}[t]{0.10\columnwidth}\raggedright
-4.5\%\strut
\end{minipage}\tabularnewline
\begin{minipage}[t]{0.07\columnwidth}\raggedright
1 KB\strut
\end{minipage} & \begin{minipage}[t]{0.12\columnwidth}\raggedright
1.63\strut
\end{minipage} & \begin{minipage}[t]{0.12\columnwidth}\raggedright
1.71\strut
\end{minipage} & \begin{minipage}[t]{0.10\columnwidth}\raggedright
1.38\strut
\end{minipage} & \begin{minipage}[t]{0.10\columnwidth}\raggedright
1.40\strut
\end{minipage} & \begin{minipage}[t]{0.09\columnwidth}\raggedright
1.64\strut
\end{minipage} & \begin{minipage}[t]{0.08\columnwidth}\raggedright
1.72\strut
\end{minipage} & \begin{minipage}[t]{0.10\columnwidth}\raggedright
-4.7\%\strut
\end{minipage}\tabularnewline
\begin{minipage}[t]{0.07\columnwidth}\raggedright
2 KB\strut
\end{minipage} & \begin{minipage}[t]{0.12\columnwidth}\raggedright
1.15\strut
\end{minipage} & \begin{minipage}[t]{0.12\columnwidth}\raggedright
1.21\strut
\end{minipage} & \begin{minipage}[t]{0.10\columnwidth}\raggedright
0.97\strut
\end{minipage} & \begin{minipage}[t]{0.10\columnwidth}\raggedright
0.97\strut
\end{minipage} & \begin{minipage}[t]{0.09\columnwidth}\raggedright
1.16\strut
\end{minipage} & \begin{minipage}[t]{0.08\columnwidth}\raggedright
1.22\strut
\end{minipage} & \begin{minipage}[t]{0.10\columnwidth}\raggedright
-5.0\%\strut
\end{minipage}\tabularnewline
\begin{minipage}[t]{0.07\columnwidth}\raggedright
4 KB\strut
\end{minipage} & \begin{minipage}[t]{0.12\columnwidth}\raggedright
0.82\strut
\end{minipage} & \begin{minipage}[t]{0.12\columnwidth}\raggedright
0.86\strut
\end{minipage} & \begin{minipage}[t]{0.10\columnwidth}\raggedright
0.68\strut
\end{minipage} & \begin{minipage}[t]{0.10\columnwidth}\raggedright
0.68\strut
\end{minipage} & \begin{minipage}[t]{0.09\columnwidth}\raggedright
0.82\strut
\end{minipage} & \begin{minipage}[t]{0.08\columnwidth}\raggedright
0.86\strut
\end{minipage} & \begin{minipage}[t]{0.10\columnwidth}\raggedright
-4.7\%\strut
\end{minipage}\tabularnewline
\bottomrule
\end{longtable}

AVLL HLDLC outperforms ExaLogLog ML at every memory point by 4--5\%
(Figure 1). The advantage is stable across the full range (0.25--4 KB),
suggesting that the fundamental tradeoff between per-register
information depth and register count has a stable equilibrium: AVLL's
5.5$\times$ register-count advantage consistently outweighs ExaLogLog's richer
per-register encoding.

\textbf{Memory-variance product.} Using the MAE-to-RMSE conversion from
§5.1 (RMSE $\approx$ MAE / 0.798), AVLL's empirical MVP at 1 KB is
\((0.01630 / 0.798)^2 \times 8{,}192 \approx 3.42\). For comparison,
ExaLogLog's empirical MVP at 1 KB is
\((0.01710 / 0.798)^2 \times 8{,}192 \approx 3.76\), consistent with its
published theoretical value of 3.78. AVLL's MVP of 3.42 is below not
only ExaLogLog's practical configuration (3.78) but its theoretical
optimum ELL(2,20) (3.67), making AVLL the most accurate cardinality
estimator per bit yet demonstrated in our comparison. We note that this
is an empirical MAE-derived figure using the Gaussian MAE/RMSE
conversion, whereas Ertl's values are asymptotic; the §5.1 validation
table confirms the conversion is accurate to within 0.3\% for these
estimators.

\begin{figure}
\centering
\includegraphics{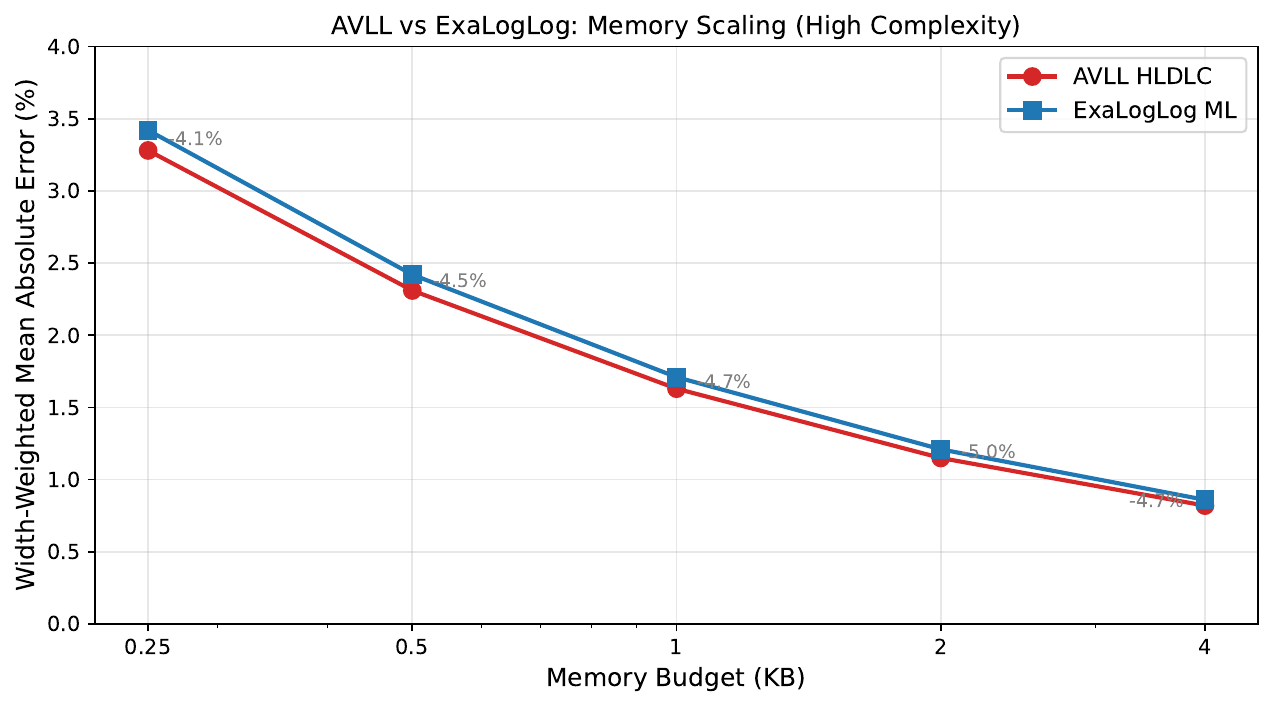}
\caption{Width-weighted mean absolute error vs memory budget (0.25--4
KB). AVLL HLDLC (red) maintains a consistent 4--5\% advantage over
ExaLogLog ML (blue) across the full range. Annotations show the relative
improvement at each point.}
\end{figure}

\hypertarget{comprehensive-1-kb-comparison-high-complexity}{%
\subsection{6.2 Comprehensive 1 KB Comparison (High
Complexity)}\label{comprehensive-1-kb-comparison-high-complexity}}

\textbf{Table 9.} All estimators at 1 KB equivalent memory,
high-complexity conditions, 128,000 instances. Each estimator uses its
best available estimation method. Sorted by WidthWtAbsErr (primary
metric).

\begin{longtable}[]{@{}lllllll@{}}
\toprule
Estimator & Registers & Bits/reg & WidthWt\% & LogWt\% & Peak\% & Victim
cache\tabularnewline
\midrule
\endhead
\textbf{AVLL HLDLC} & \textbf{1,408} & \textbf{5.82} & \textbf{1.63} &
\textbf{1.38} & \textbf{1.64} & \textbf{No}\tabularnewline
ExaLogLog ML & 256 & 32 & 1.71 & 1.40 & 1.72 & No\tabularnewline
HLL-TailCut4 & 2,048 & 4 & 1.83 & 1.61 & 3.10 & No\tabularnewline
HyperLogLogLog & 2,048 & 3+vc & 1.83 & 1.62 & 3.10 & Yes
(IntHashMap)\tabularnewline
Apache HLL4 & 2,048 & 4+vc & 1.83 & 1.62 & 3.10 & Yes
(HashMap)\tabularnewline
HyperTwoBits & 4,096 & 2 & 1.90 & 3.37 & 100$\dagger$ & No\tabularnewline
UltraLogLog (FGRA) & 1,024 & 8 & 1.95 & 1.74 & 1.96 & No\tabularnewline
LogLog30 (HLL) & 1,280 & 6.4 & 2.32 & 2.06 & 3.47 & No\tabularnewline
\bottomrule
\end{longtable}

Victim cache = unbounded auxiliary memory maps required for register
overflow; all other estimators use only their fixed-size register
arrays. $\dagger$HyperTwoBits diverges at low cardinality (100\% error at
cardinality 1, 24.7\% at cardinality 10); above cardinality 1,000 its
peak is 2.86\%.

AVLL is the most accurate estimator at 1 KB, followed by ExaLogLog.
HLL-TailCut4, HyperLogLogLog, and Apache HLL4 cluster at 1.83\%
width-weighted error --- all three use 2,048 registers with the same
harmonic-mean estimation pipeline, producing identical or near-identical
estimates at every cardinality (HLL-TailCut4 and HyperLogLogLog are
lossless re-encodings of the same 6-bit register state; Apache HLL4 uses
an equivalent 4-bit offset representation with a victim cache that
reproduces the same estimates below offset saturation). HyperTwoBits
achieves competitive width-weighted error (1.90\%) despite using only 2
bits per register, but it diverges at low cardinality (100\% error at
cardinality 1, marked with $\dagger$), and its 3.37\% log-weighted error
indicates poor low-cardinality accuracy.

UltraLogLog with the FGRA estimator {[}5{]} achieves 1.95\%
width-weighted error at 1,024 registers --- competitive given its
simpler 8-bit-per-register encoding and closed-form estimator. LogLog30
with the standard HLL harmonic mean achieves 2.32\% width-weighted error
with a 3.47\% peak error.

\hypertarget{low-complexity-robustness-lc100m}{%
\subsection{6.3 Low-Complexity Robustness
(LC100M)}\label{low-complexity-robustness-lc100m}}

\textbf{Table 10.} Width-weighted mean absolute error under
low-complexity conditions (LC100M) (100-million-element pool,
\textasciitilde87 million observed distinct values after 400 million
additions with skewed duplicates), 32,000 independent instances at 1 KB.

\begin{longtable}[]{@{}llll@{}}
\toprule
Estimator & WidthWt\% (LC) & WidthWt\% (HC) & \(\Delta\) LC vs
HC\tabularnewline
\midrule
\endhead
\textbf{AVLL HLDLC} & \textbf{1.63} & \textbf{1.63} &
\textbf{0.0\%}\tabularnewline
ExaLogLog ML & 1.72 & 1.71 & +0.6\%\tabularnewline
HLL-TailCut4 & 1.83 & 1.83 & 0.0\%\tabularnewline
HyperLogLogLog & 1.83 & 1.83 & 0.0\%\tabularnewline
Apache HLL4 & 1.83 & 1.83 & 0.0\%\tabularnewline
HyperTwoBits & 1.71 & 1.90 & -10.0\%\tabularnewline
UltraLogLog (FGRA) & 1.95 & 1.95 & 0.0\%\tabularnewline
LogLog30 (HLL) & 2.32 & 2.32 & 0.0\%\tabularnewline
\bottomrule
\end{longtable}

AVLL shows no accuracy degradation under low-complexity data (LC WidthWt
= HC WidthWt = 1.63\%). This is a critical result because non-idempotent
estimators (including AVLL, which uses tier promotion) could in
principle be vulnerable to systematic bias from duplicate elements. The
LC100M test confirms that AVLL's HLDLC estimator is robust to
duplication at practical cardinalities. The significance of this null
result is discussed in §7.3.

All estimators show negligible LC vs HC divergence at 1 KB (the full
per-cardinality LC error profiles confirm that the error shapes are not
distorted by duplication, not just the aggregate values). HyperTwoBits
is the exception: its LC accuracy actually \emph{improves} relative to
HC (1.71\% vs 1.90\%), likely because the LinearCounting-like estimation
formula that HTB uses is naturally robust to duplication.

\hypertarget{per-cardinality-error-curves}{%
\subsection{6.4 Per-Cardinality Error
Curves}\label{per-cardinality-error-curves}}

Figure 2 shows the per-cardinality absolute error for AVLL HLDLC vs
ExaLogLog ML at 1 KB. AVLL's error is lower at nearly all cardinalities,
with the advantage most pronounced at high cardinality where the HLDLC
blend of LC-based and harmonic-mean-based estimators achieves superior
bias cancellation. At very low cardinality (below \textasciitilde20),
ExaLogLog's ML estimator is slightly more accurate, though both
estimators' absolute errors are small in this range.

\begin{figure}
\centering
\includegraphics{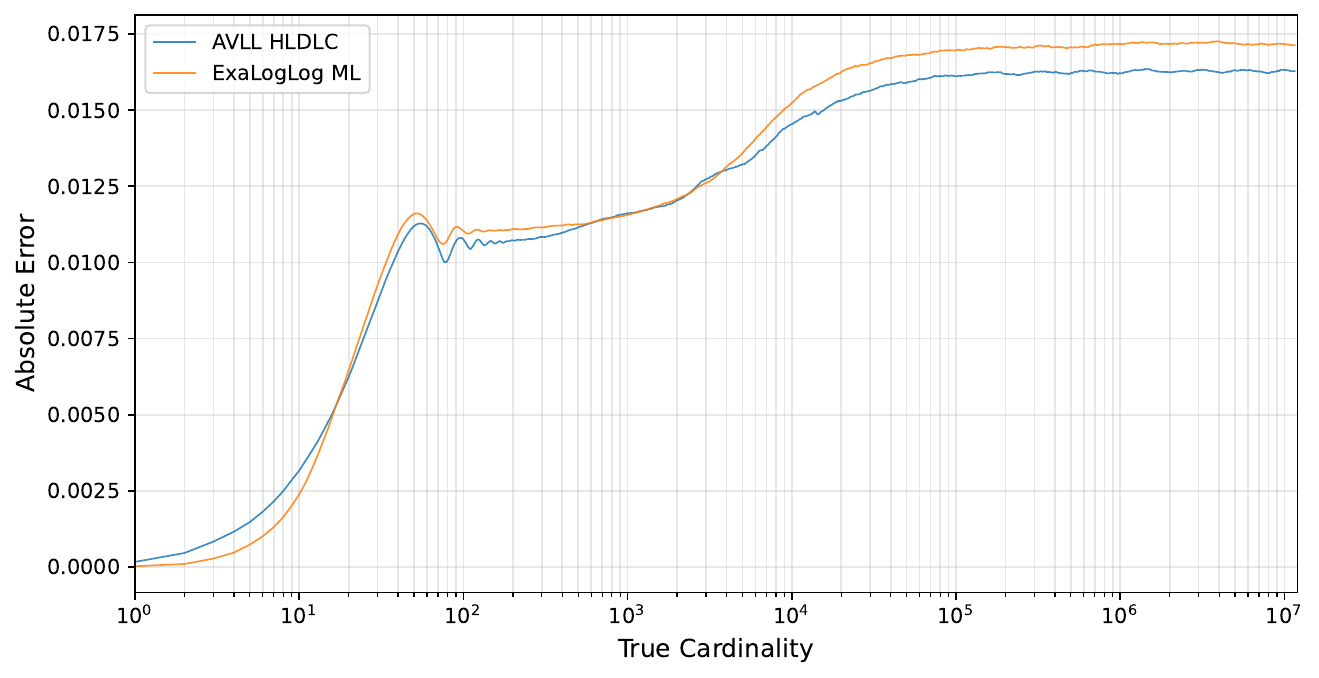}
\caption{Per-cardinality absolute error at 1 KB: AVLL HLDLC vs ExaLogLog
ML. AVLL (blue) maintains lower error at nearly all cardinalities
(ExaLogLog is slightly more accurate below \textasciitilde20). Both
curves show error rising from the low-cardinality regime into the
asymptotic LogLog regime (the transition from LC-dominated to
harmonic-mean-dominated estimation), then plateauing. AVLL's asymptotic
error ceiling is lower.}
\end{figure}

Figure 3 shows the comprehensive comparison of all calibrated estimators
at 1 KB. The HLL-family estimators (HTC4, HLLL, HLL4) show
characteristic LC-to-LogLog transition spikes around cardinality
1,000--10,000 that AVLL and ExaLogLog avoid through smoother estimation
pipelines. HyperTwoBits exhibits a large error spike at very low
cardinality (visible at the left edge) that contributes to its high peak
error (Table 9).

\begin{figure}
\centering
\includegraphics{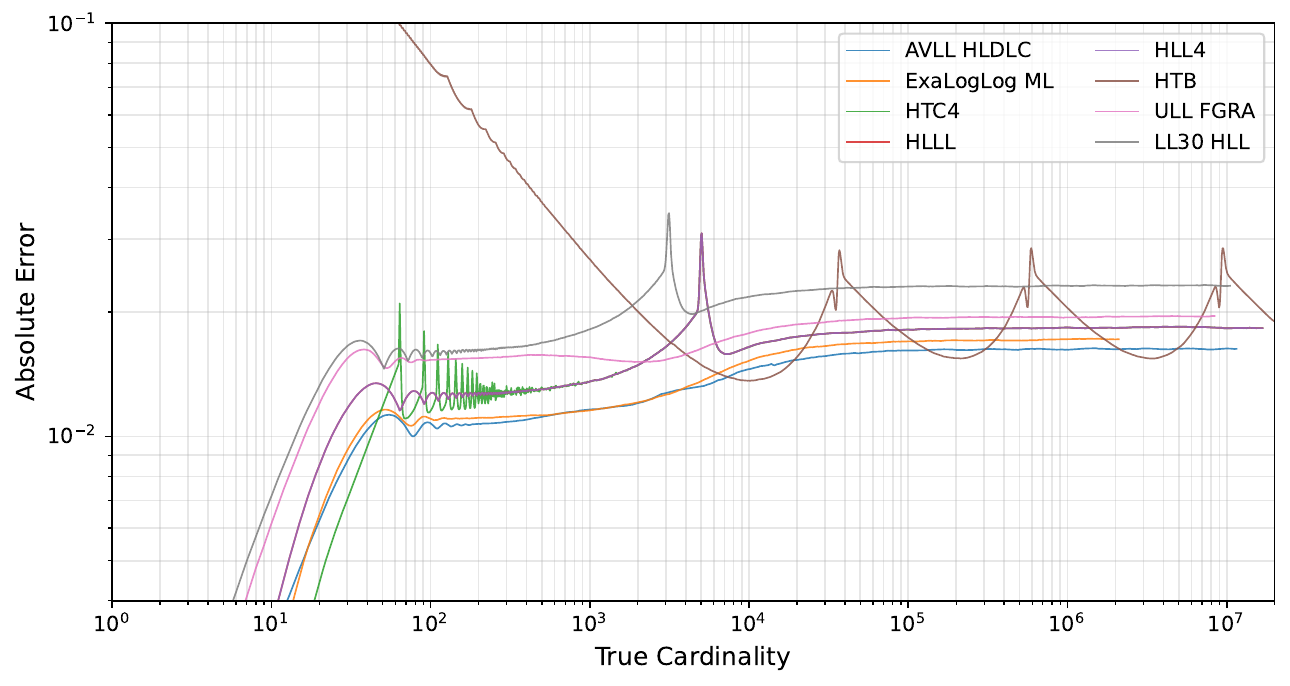}
\caption{Per-cardinality absolute error at 1 KB (log scale):
comprehensive comparison. AVLL HLDLC (blue) is the lowest curve at
nearly all cardinalities. ExaLogLog ML (orange) is second. HLL-family
estimators (HTC4, HLLL, HLL4) show transition spikes around cardinality
1,000--10,000. HyperTwoBits (brown) diverges at low cardinality,
dropping from 100\% error at cardinality 1. X-axis ranges differ across
estimators because the HC protocol runs each to 8,192 \(\times\)
\emph{B}, and register counts vary at equal memory (Section 5.2).}
\end{figure}

Figure 4 shows the per-cardinality error of each HLDLC component
estimator at 1 KB, illustrating how the components complement each other
across the cardinality range. SBS (blue) is accurate at low cardinality
but diverges above \textasciitilde6$\times$\emph{B}; Mean+H (orange) diverges
at low cardinality but converges at high; HC (green) requires multiple
populated tiers and converges at high cardinality; DLC (red) is accurate
everywhere but with periodic oscillation; LDLC (purple) and Hybrid+2
(brown) are smooth across the full range; HLDLC (pink) is the lowest,
combining the best of both.

\begin{figure}
\centering
\includegraphics{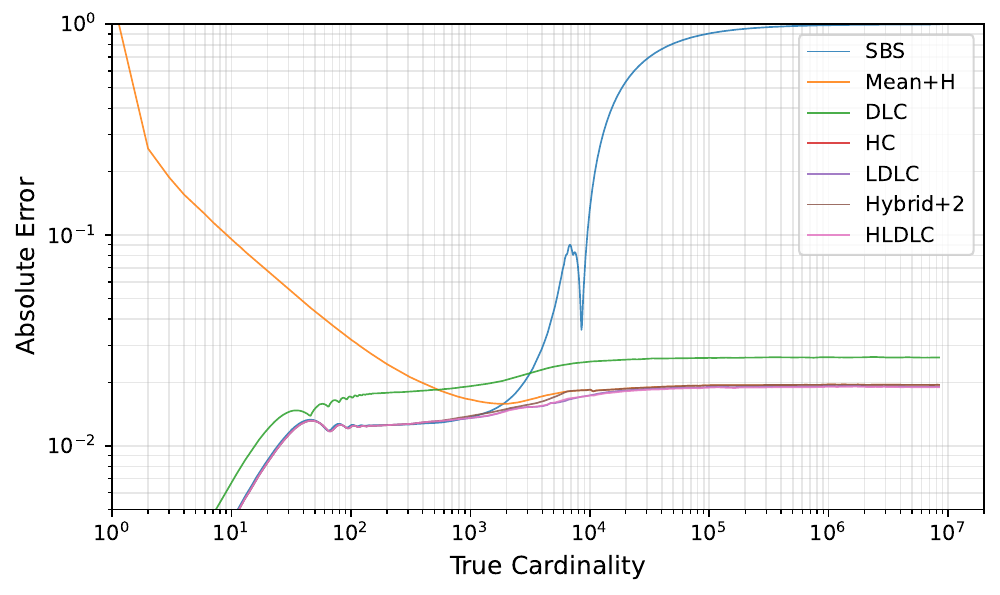}
\caption{HLDLC component accuracy at 1 KB (log scale). Each component is
accurate in a different cardinality range. SBS (blue) is accurate at low
cardinality but diverges above \textasciitilde6\(\times\)\emph{B};
Mean+H (orange) diverges at low cardinality but converges at high; HC
(green) converges at high cardinality as tiers populate; DLC (red) works
everywhere but oscillates; LDLC (purple) and Hybrid+2 (brown) are smooth
full-range estimators; HLDLC (pink) achieves the lowest error through
complementary bias cancellation.}
\end{figure}

\hypertarget{throughput}{%
\subsection{6.5 Throughput}\label{throughput}}

\textbf{Table 11.} Insertion throughput in million adds/second on a
64-core cluster node (2 KB equivalent memory). Three scenarios test
different memory-pressure regimes (protocols in §5.3). Sorted by
sim=4,096 performance. Abbreviated names: LL6 = LogLog30 (HLL), HTC4 =
HLL-TailCut4, HLLL = HyperLogLogLog, HLL4 = Apache HLL4, ULL =
UltraLogLog, HTB = HyperTwoBits. DLL4 and UDLL are DynamicLogLog
variants benchmarked for throughput comparison; their accuracy is
reported in {[}6{]}.

\begin{longtable}[]{@{}llllll@{}}
\toprule
Estimator & ALU & sim=2,048 & sim=4,096 & Early exit & Slowdown
ALU$\rightarrow$4K\tabularnewline
\midrule
\endhead
HTB & 19,611 & 14,494 & 11,807 & Yes & 1.7$\times$\tabularnewline
DLL4 & 21,272 & 15,430 & 10,680 & Yes & 2.0$\times$\tabularnewline
HTC4 & 15,447 & 10,996 & 7,881 & Yes & 2.0$\times$\tabularnewline
UDLL & 15,554 & 9,861 & 5,482 & Yes & 2.8$\times$\tabularnewline
HLLL & 13,461 & 8,135 & 4,699 & Partial & 2.9$\times$\tabularnewline
\textbf{AVLL} & \textbf{14,681} & \textbf{8,772} & \textbf{4,360} &
\textbf{Yes} & \textbf{3.4$\times$}\tabularnewline
LL6 (HLL) & 11,218 & 2,599 & 1,712 & No & 6.6$\times$\tabularnewline
HLL4 (Apache) & 8,664 & 2,325 & 1,649 & No & 5.3$\times$\tabularnewline
ExaLogLog & 13,801 & 1,960 & 1,609 & No & 8.6$\times$\tabularnewline
ULL & 10,820 & 2,324 & 1,587 & No & 6.8$\times$\tabularnewline
\bottomrule
\end{longtable}

The results reveal a sharp divide determined by the \textbf{early exit
mechanism} (Section 3.4).

Estimators with early exit (DLL-family: DLL4, HTC4, UDLL, AVLL, and the
related HTB) maintain high throughput even under extreme cache pressure
because at high cardinality, the vast majority of elements are rejected
after a single hash and comparison --- no register memory is touched
(Table 12: \textasciitilde96\% rejection at 20M cardinality, increasing
monotonically with cardinality). The sketch could be entirely out of
cache and it would not matter for rejected elements. HTB shows minimal
slowdown from ALU to sim=4,096 (1.7$\times$) because its early exit rate is the
same regardless of how many sketches are in flight.

Estimators without early exit (ExaLogLog, ULL, LL6, HLL4) must access
register memory on every element. Every \texttt{add()} reads a register
before it can determine whether an update is needed, incurring a cache
miss when the sketch is out of cache. ExaLogLog drops from 13,801 to
1,609 M/s --- an 8.6$\times$ slowdown --- because it must read the register
value to compute the delta before it can decide to reject an element. In
contrast, DLL-family estimators compare the hash against a single global
threshold (\texttt{globalNLZ}) that stays in a register or L1 cache,
rejecting the vast majority of elements without touching any register
memory.

Under cache pressure, AVLL is 2.7--4.5$\times$ faster than ExaLogLog (4,360 vs
1,609 M/s at sim=4,096; 8,772 vs 1,960 at sim=2,048). AVLL's 3.4$\times$
slowdown from ALU to sim=4,096 is due to the rare register updates that
do occur at high cardinality, plus the division overhead when those
updates require \texttt{Long.divideUnsigned} --- but these updates are
rare enough that the overall throughput remains competitive.

Under ALU pressure (all sketches in L1/L2), AVLL achieves 14,681 M/s ---
comparable to ExaLogLog's 13,801 M/s (within the 10\% run-to-run
variance). AVLL's early exit rejects most elements before the expensive
\texttt{Long.remainderUnsigned}, but every element still pays a hash and
comparison; ExaLogLog reads a register on every element but uses a
simple power-of-two array access. When the register array fits in cache,
these per-element costs are comparable.

\hypertarget{throughput-scaling}{%
\subsection{6.6 Throughput Scaling}\label{throughput-scaling}}

Table 11 captures throughput at a single memory budget (2 KB) with two
levels of cache pressure. To understand how these relationships change
across deployment regimes, we vary three parameters independently:
estimator memory (Figure 5), cache pressure (Figure 6), and stream
cardinality (Figure 7). All measurements use 128 threads on a 64-core
node with exclusive scheduling.

\hypertarget{memory-scaling}{%
\subsubsection{Memory scaling}\label{memory-scaling}}

\begin{figure}
\centering
\includegraphics[width=1\textwidth,height=\textheight]{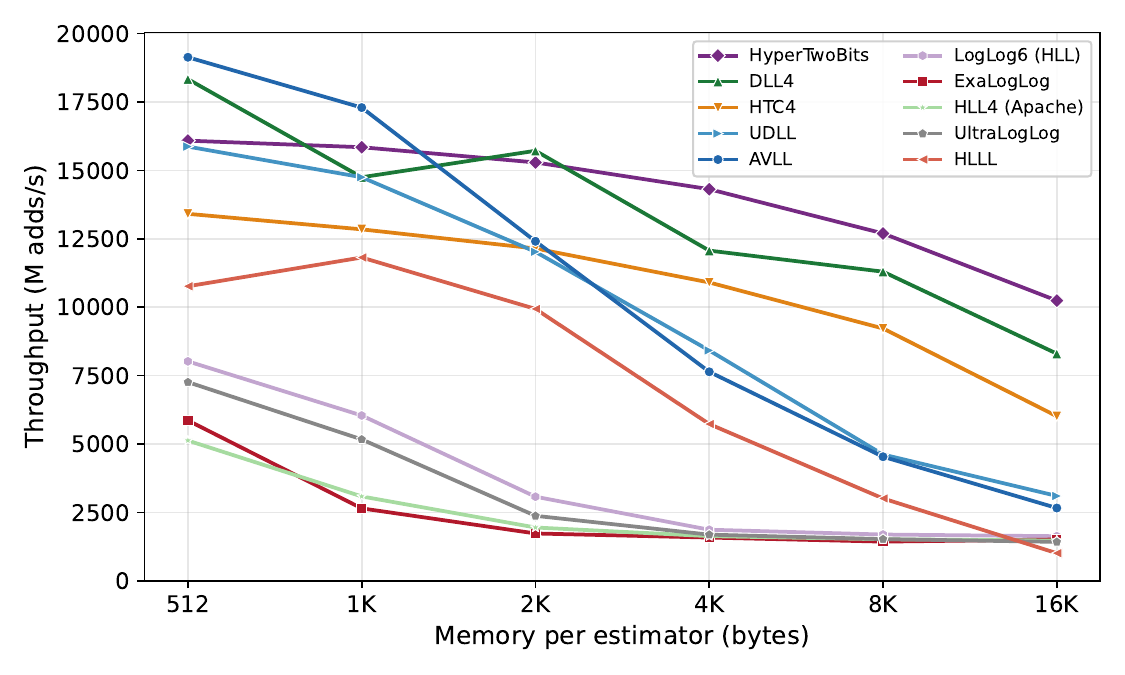}
\caption{Throughput vs.~estimator memory at fixed cache pressure
(sim=2,048, card=20M). Each estimator receives an equal memory budget;
bucket counts vary by encoding density. Early-exit types (DLL4, HTB,
HTC4) degrade gracefully. Types without early exit (ExaLogLog, ULL, LL6,
HLL4) converge to a shared bandwidth floor near 1,500 M/s. HLLL degrades
steeply despite having early exit due to IntHashMap-based exception
storage.}
\end{figure}

Figure 5 reveals three regimes determined by the interaction between
early exit and per-access cost.

At the top, \textbf{HTB, DLL4, and HTC4} maintain high throughput across
all memory budgets. HTB drops only 36\% from 512 bytes to 16 KB (16,086
$\rightarrow$ 10,241 M/s) because its 2-bit registers keep per-access cost minimal,
and the early exit mechanism ensures that the vast majority of elements
never touch register memory regardless of estimator size. DLL4 (4-bit
registers) and HTC4 follow similar trajectories.

In the middle band, \textbf{AVLL and UDLL} start fast --- AVLL leads all
types at 512 bytes (19,134 M/s) --- but degrade more steeply (AVLL drops
86\% to 2,661 at 16 KB). Both have early exit, but larger memory budgets
mean more registers, each receiving fewer updates, which lowers the
global NLZ floor and reduces the early exit rate. When early exit fails,
AVLL pays 10-19 cycles for each \texttt{Long.divideUnsigned} in
arithmetic register access, while DLL4 pays \textasciitilde3 cycles for
a nibble lookup. At small memory the division rarely triggers; at large
memory it dominates.

\textbf{HLLL} degrades the most steeply of any early-exit type (10,765 $\rightarrow$
1,010 M/s, 91\%). HLLL stores registers as compact 3-bit offsets from a
global base, with an IntHashMap-based exception map for values that
overflow the offset range. Its early exit mechanism is the same as
DLL-family --- a global minimum check --- but when early exit fails,
every access must probe the exception map before it can determine
whether the register is in the compressed array. This map lookup is far
more expensive than a byte or nibble access and pollutes additional
cache lines.

At the bottom, \textbf{ExaLogLog, ULL, LL6, and HLL4} converge to
approximately 1,500 M/s at 16 KB. Without early exit, every
\texttt{add()} reads a register, incurring a cache miss when the
estimator is out of cache. At large memory, all four are
memory-bandwidth-bound and their per-access logic differences are hidden
by cache latency. ExaLogLog's trajectory is shallower than the
early-exit types (5,861 $\rightarrow$ 1,510 M/s, a 74\% drop) because it lacks early
exit --- it accesses a register on every insertion regardless of memory
size, so scaling up the register array adds only cache-miss cost, not
the loss of an early-exit filter that the other types depend on.

\hypertarget{cache-pressure-scaling}{%
\subsubsection{Cache pressure scaling}\label{cache-pressure-scaling}}

\begin{figure}
\centering
\includegraphics[width=1\textwidth,height=\textheight]{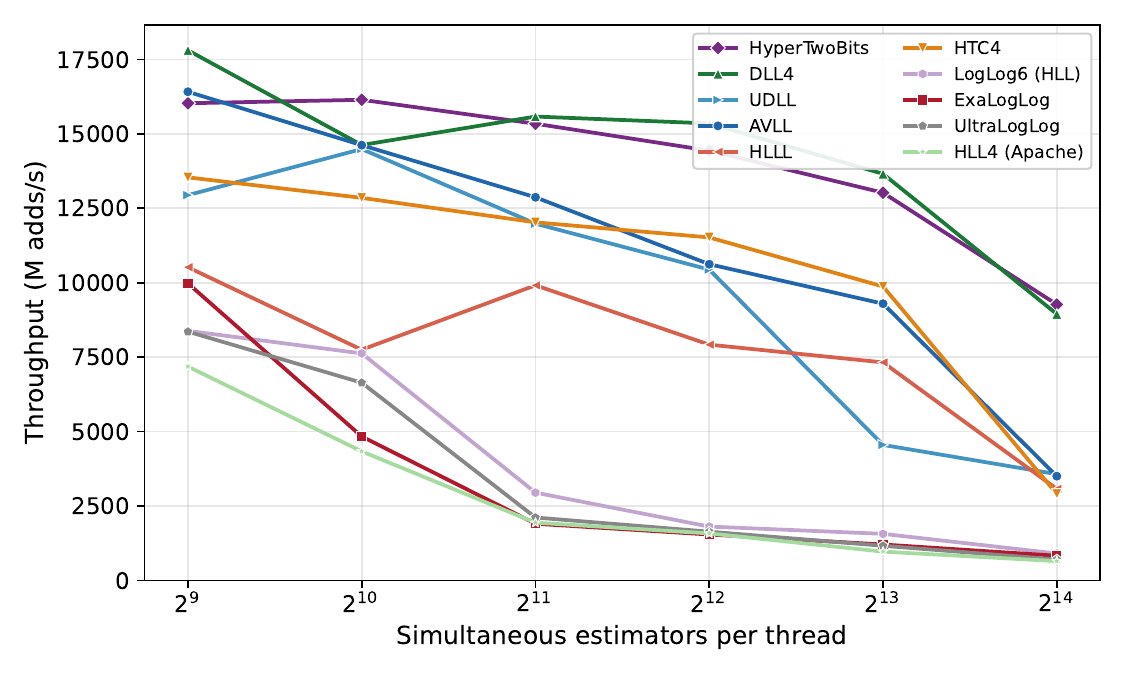}
\caption{Throughput vs.~simultaneous estimators per thread at fixed
memory (mem=2 KB, card=20M). HTB and DLL4 maintain high throughput even
at 16,384 simultaneous estimators. HTC4 and HLLL collapse at extreme
pressure despite having early exit. ExaLogLog drops from 9,978 to 826
M/s.}
\end{figure}

Figure 6 isolates the cache pressure effect by varying the number of
simultaneously active estimators while holding estimator size constant
at 2 KB. The total working set ranges from 128 MB (sim=512) to 4 GB
(sim=16,384).

\textbf{DLL4} is the most cache-resistant type tested, dropping 50\%
across a 32$\times$ increase in estimator count (17,807 $\rightarrow$ 8,945 M/s) --- the
fastest or tied with HTB at every pressure level. Its 4-bit registers
and early exit mechanism keep throughput high even with over 2 million
simultaneous estimators.

\textbf{AVLL} degrades 79\% (16,414 $\rightarrow$ 3,499 M/s) but remains the fastest
of the high-accuracy estimators at every pressure level and 4.2$\times$ faster
than ExaLogLog at sim=16,384. The steeper decline relative to DLL4
reflects the division cost: when early exit fails at extreme pressure,
AVLL's \texttt{Long.divideUnsigned} is more expensive than DLL4's nibble
access.

\textbf{ExaLogLog} exhibits the steepest decline (92\%, from 9,978 to
826 M/s). At sim=512 the working set is small enough that some
estimators remain in L3 cache between accesses; by sim=16,384,
essentially every \texttt{add()} is a cache miss. Without early exit,
the miss rate translates directly into throughput loss. ULL, LL6, and
HLL4 follow similar trajectories, all dropping below 1,000 M/s.

\hypertarget{cardinality-scaling}{%
\subsubsection{Cardinality scaling}\label{cardinality-scaling}}

\begin{figure}
\centering
\includegraphics[width=1\textwidth,height=\textheight]{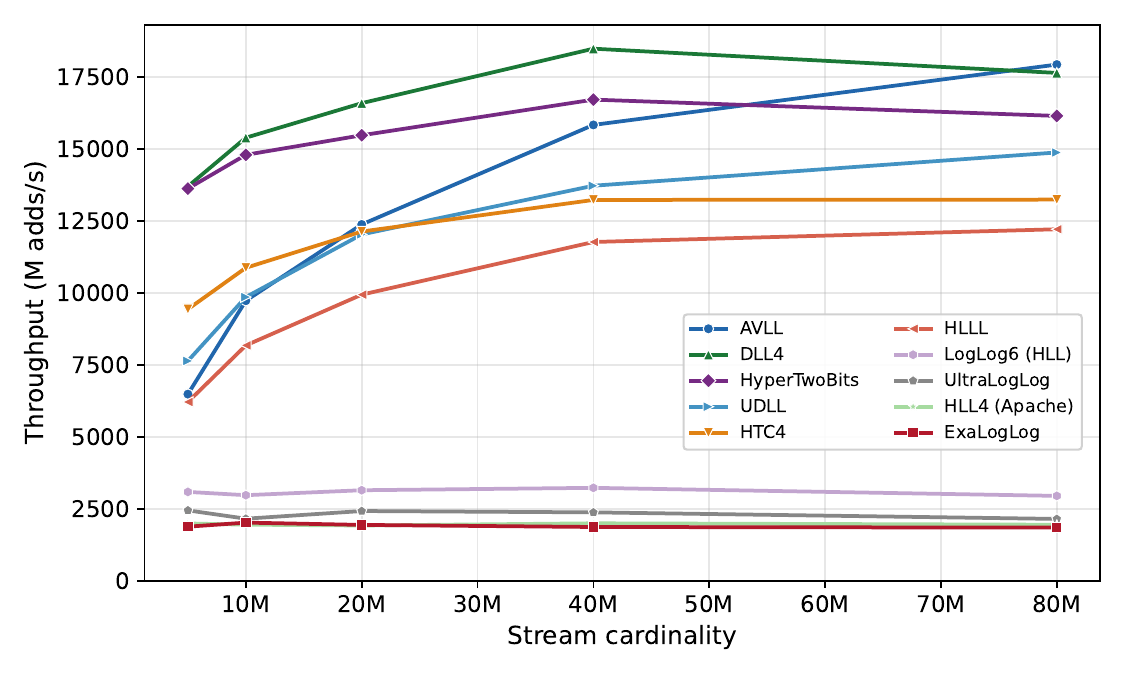}
\caption{Throughput vs.~stream cardinality at fixed memory and cache
pressure (mem=2 KB, sim=2,048). All early-exit types accelerate with
cardinality. AVLL catches DLL4 at 80M (equal within run-to-run variance)
as the early exit becomes dominant. ExaLogLog and non-DLL types are
flat.}
\end{figure}

Figure 7 reveals a throughput dimension not visible in fixed-cardinality
benchmarks: types with early exit \emph{accelerate} as stream
cardinality increases.

\textbf{AVLL} speeds up 177\% from 5M to 80M cardinality (6,481 $\rightarrow$ 17,922
M/s). At low cardinality, the global NLZ floor is low --- most hash
values can still improve some register, so early exit rejects a smaller
fraction of elements and more additions pay the full arithmetic
register-access cost. As cardinality grows and registers saturate, the
global NLZ rises, early exit fires on a larger fraction of elements, and
throughput climbs. At 80M, AVLL matches DLL4 within run-to-run variance
(17,922 vs 17,632 M/s, \textasciitilde1.6\% margin) --- the division
overhead that makes AVLL slower at low cardinality becomes negligible
when the vast majority of elements are rejected before reaching the
register-access path. \textbf{UDLL} shows a similar pattern (+95\%,
7,634 $\rightarrow$ 14,871 M/s). \textbf{HLLL}, \textbf{HTC4}, and \textbf{HTB} also
accelerate, confirming that the effect is universal across early-exit
types.

\textbf{DLL4} improves a more modest 29\% (13,693 $\rightarrow$ 17,632 M/s). Its
4-bit registers and cheaper per-access cost mean early exit is already
highly effective at 5M --- less room to improve.

\textbf{ExaLogLog} is flat across the entire range (1,880 to 1,847 M/s,
-2\%). Without early exit, every element touches register memory
regardless of how saturated the registers are. \textbf{ULL},
\textbf{LL6}, and \textbf{HLL4} are similarly flat.

This cardinality effect means AVLL's speed advantage over ExaLogLog
grows further in high-cardinality streaming applications. The
AVLL-to-ExaLogLog throughput ratio rises from 3.4$\times$ at 5M to 9.7$\times$ at 80M;
at higher cardinalities the trend continues because AVLL's early exit
rate is monotonically increasing in cardinality while ExaLogLog's
per-element cost is constant.

\hypertarget{register-access-rates}{%
\subsubsection{Register access rates}\label{register-access-rates}}

\textbf{Table 12.} Fraction of \texttt{add()} calls that result in a
register read (passed early exit) and a register write (actual update),
at 2 KB equal memory and 20 million cardinality. Writes/bucket is the
total register updates divided by bucket count, measuring per-register
write pressure. Sorted by writes/bucket.

\begin{longtable}[]{@{}lllllll@{}}
\toprule
Estimator & Buckets & Read \% & Write \% & Writes/bucket & Early exit &
History bits\tabularnewline
\midrule
\endhead
HTB & 8,192 & 0.520 & 0.125 & 3.0 & Yes (T, T+4) & 0\tabularnewline
DLL4 & 4,096 & 0.965 & 0.139 & 6.8 & Yes & 0\tabularnewline
HLL4 & 4,096 & 100 & 0.139 & 6.8 & No & 0\tabularnewline
HLLL & 4,096 & 0.982 & 0.139 & 6.8 & Partial & 0\tabularnewline
HTC4 & 4,096 & 1.455 & 0.139 & 6.8 & Yes & 0\tabularnewline
LL6 (HLL) & 2,560 & 100 & 0.092 & 7.2 & No & 0\tabularnewline
\textbf{AVLL} & \textbf{2,816} & \textbf{4.042} & \textbf{0.151} &
\textbf{10.7} & \textbf{Yes} & \textbf{2}\tabularnewline
UDLL & 2,560 & 3.632 & 0.139 & 10.9 & Yes & 2\tabularnewline
ULL & 2,048 & 100 & 0.113 & 11.1 & No & 2\tabularnewline
ExaLogLog & 512 & 100 & 0.129 & 50.6 & No & 24\tabularnewline
\bottomrule
\end{longtable}

The read rate cleanly separates the early-exit types from the
non-early-exit types. Types with early exit (HTB, DLL4, HLLL, HTC4,
AVLL, UDLL) read a register on fewer than 6\% of additions at 20M
cardinality; types without early exit (ExaLogLog, ULL, LL6, HLL4) read a
register on every single addition. This is the dominant factor in
throughput under cache pressure: a register read on a cache-cold sketch
is a guaranteed cache miss.

Among the early-exit types, HTB achieves the lowest read rate (0.52\%)
because its threshold advances in steps of 4 rather than 1 --- each
advance raises the rejection bar by 4 NLZ values, so fewer hashes clear
the threshold. This explains why HTB matches DLL4's throughput at
extreme cache pressure (§6.6) despite using two separate bit-arrays per
register access.

The writes/bucket column reveals a second, subtler factor:
\textbf{per-register write pressure scales with history depth}. DLL4,
HLL4, HLLL, and HTC4 all share the same bucket count (4,096) and all
store only the maximum NLZ --- their writes/bucket values are identical
(6.8), confirming that write rate is determined purely by
birthday-problem dynamics when no history is recorded. AVLL, UDLL, and
ULL store 2-bit sub-NLZ history, which introduces additional states that
can be improved without advancing the NLZ, raising per-bucket writes to
10.7--11.1. ExaLogLog's 24-bit observation history raises per-bucket
writes to 50.6 --- an element can update a register's history bits even
when its NLZ is \emph{lower} than the current maximum. Each of these
extra writes is a cache miss under deployment conditions.

The combination of read rate and per-bucket write pressure explains the
full throughput hierarchy. ExaLogLog pays twice: 100\% read rate (no
early exit) and 7.4$\times$ more writes per bucket than the no-history group
(50.6 vs 6.8 --- history amplification from its 24-bit observation
record). AVLL pays a moderate history tax (1.6$\times$ more writes/bucket than
DLL4) but compensates with 96\% early exit, making register access rare.
HTB achieves the best of both: the lowest read rate (0.52\%) and the
lowest per-bucket writes (3.0), at the cost of accuracy (Table 9).

\hypertarget{discussion}{%
\section{7. Discussion}\label{discussion}}

\hypertarget{the-design-principle-density-beats-richness}{%
\subsection{7.1 The Design Principle: Density Beats
Richness}\label{the-design-principle-density-beats-richness}}

The central question is not how many registers each estimator has, but
how much information about cardinality each estimator extracts per bit
of memory. All estimators in our comparison operate on the same memory
budget; they differ in how they encode and decode observations from
their data structures.

ExaLogLog dedicates 32 bits to each register, encoding sub-NLZ
fractional bits and 24 bits of observation history per register. Its ML
estimator is Fisher-information-optimal for this encoding --- it
extracts the theoretical maximum information from each register's state.
At 1 KB, this yields 256 registers with very high per-register
information content.

AVLL dedicates 5.82 bits per register, encoding 2-bit sub-NLZ history
and an NLZ exponent via arithmetic packing. Its HLDLC estimator is a
closed-form blend rather than an optimal ML solver, leaving some
per-register information unused. At 1 KB, this yields 1,408 registers
with lower per-register information content.

The empirical result is clear: AVLL wins by 4--5\% at every memory
point. The additional statistical independence from 5.5$\times$ more
independent hash observations outweighs the information lost to simpler
per-register encoding. This margin is stable across the tested memory
range (0.25--4 KB) because the architectural parameters are
scale-invariant: AVLL always uses 5.82 bits per register, ExaLogLog
always uses 32. The design principle --- that for LogLog-family
estimators, the most productive use of memory is often more registers
rather than richer registers, provided the estimator is sophisticated
enough to extract the available information --- is the central result of
this paper.

\hypertarget{the-price-of-density}{%
\subsection{7.2 The Price of Density}\label{the-price-of-density}}

Arithmetic encoding's primary cost is the division operation in register
access. Unlike HLL's simple byte-array access or DLL4's 4-bit nibble
extraction, AVLL requires \texttt{Long.divideUnsigned} and
\texttt{Long.remainderUnsigned} for each register read and write. On
modern x86 processors, each 64-bit unsigned division takes approximately
10-19 cycles --- significantly more than the 1--3 cycles for a
byte-array access. However, the early exit mask ensures that the vast
majority of elements at high cardinality never reach the register-access
path (§3.4), so the division cost is amortized over only the rare
updates.

AVLL is not idempotent (§7.3) and its register-wise merge inherits
slightly more overflow from independent tier promotion than a
single-stream instance (§7.5).

\hypertarget{when-to-choose-avll}{%
\subsection{7.3 When to Choose AVLL}\label{when-to-choose-avll}}

Across every tested memory budget, AVLL is simultaneously the most
accurate (§6.1--6.2) and, under deployment-scale cache pressure, the
fastest of the high-accuracy estimators (§6.5). The one principled
reason to prefer ExaLogLog or UltraLogLog is strict insert idempotency.
Idempotency matters chiefly because a non-idempotent sketch could, in
principle, be biased by duplicate elements --- and Table 10 tests
exactly that failure mode, using a 100-million-element pool under heavy
duplication (\textasciitilde87 million observed distinct values),
measuring zero accuracy degradation. The practical scope of the
limitation is therefore confined to merge-only workflows in which the
same sketch is updated from overlapping data sources, or when the sketch
is used to calculate metrics other than cardinality, such as set
similarity.

\hypertarget{theoretical-vs.-empirical-optimization}{%
\subsection{7.4 Theoretical vs.~Empirical
Optimization}\label{theoretical-vs.-empirical-optimization}}

ExaLogLog's ML estimator is the theoretical optimum for extracting
information from fixed-width LogLog registers, derived from
Fisher-information analysis with provable variance bounds. AVLL's HLDLC
is a heuristically derived pipeline: its four components and
approximately 100 correction constants were fitted through massive-scale
Monte Carlo simulation (up to 512,000 independent instances) rather than
proved optimal through closed-form analysis. A formal proof of HLDLC's
variance behavior over arithmetic-encoded registers with correlated tier
promotions is, to our knowledge, intractable.

This distinction is important. ExaLogLog's MVP of 3.67 (ELL(2,20)) is a
\emph{theoretical} bound --- it is the best achievable result given its
register encoding and an information-optimal estimator. AVLL's MVP of
approximately 3.4 is an \emph{empirical} observation: it holds across
128,000 independent instances at five memory points, cross-validated
against ExaLogLog's own theoretical predictions (§5.1), but it is not
proved to be the best achievable result for AVLL's encoding. If a future
ML estimator were developed for base-56 arithmetic registers, the
accuracy gap may widen further --- though the per-register information
content is lower than ExaLogLog's, the 5.5$\times$ register-count advantage
provides substantially more independent observations for any estimator
to exploit.

The practical implication is that AVLL's accuracy advantage is robust
--- demonstrated across a wide range of memory budgets, data
distributions, and cardinality regimes --- but its optimality is not
guaranteed. ExaLogLog's theoretical proofs characterize average-case
behavior, not worst-case guarantees; the principled reason to prefer it
is idempotency (§7.3).

\hypertarget{limitations}{%
\subsection{7.5 Limitations}\label{limitations}}

\textbf{Division overhead.} AVLL's register access requires 64-bit
unsigned division (\textasciitilde10-19 cycles), slower per access than
fixed-width schemes (§7.2). Early exit mitigates this at high
cardinality; at low cardinality every element pays the cost.

\textbf{Merge overflow.} Independent tier promotion across merged
instances causes slightly more overflow than single-stream processing,
shared with all DLL-family estimators {[}6{]}.

\hypertarget{conclusion}{%
\section{8. Conclusion}\label{conclusion}}

Arithmetic Variable LogLog advances the state of the art for cardinality
estimation on three fronts.

\textbf{Record accuracy per bit.} By packing 5.5$\times$ more registers than
ExaLogLog into the same memory through base-56 arithmetic encoding, AVLL
achieves 4--5\% lower width-weighted error at every tested size (0.25--4
KB) and an empirical MVP of approximately 3.4 --- the lowest yet
measured, and below ExaLogLog's theoretical optimum. At 1 KB: 1.63\% vs
1.71\%.

\textbf{Speed.} Early exit filters the vast majority of elements before
any register memory is touched, yielding 2.7--4.5$\times$ ExaLogLog's
throughput when thousands of sketches compete for cache.

\textbf{No register-width cardinality ceiling.} Relative-NLZ encoding
with a sliding floor makes AVLL's memory cost additive in cardinality
(O(\emph{B} + log log \emph{C})) rather than multiplicative, removing
the register-width ceiling that constrains fixed-register schemes. The
maximum representable cardinality is bounded by the hash width, not by
register encoding.

These results point to a consistent design principle: density beats
richness. Given a fixed memory budget and an estimator sophisticated
enough to use the registers, more independent hash observations outweigh
richer per-register encodings at every point we measured. The
DLL/UDLL/AVLL progression illustrates this at increasing encoding
efficiency. DLL4 packs 8 registers per 32-bit word with 4-bit fixed
slots; UDLL fits 5 per word with 6-bit slots; AVLL fits 11 per 64-bit
word with 56-ary arithmetic slots. Each step trades per-access
simplicity for per-byte density, and each step lowers error.

\hypertarget{availability}{%
\section{Availability}\label{availability}}

AVLL is implemented as the class \texttt{ArithmeticVariableLogLog} in
the BBTools bioinformatics suite {[}3{]}, available at
\url{https://bbmap.org}. BBTools is free and open-source. For
\emph{k}-mer cardinality estimation, use \texttt{loglog.sh} with the
flag \texttt{loglogtype=avll}.

\hypertarget{acknowledgements}{%
\section{Acknowledgements}\label{acknowledgements}}

The author thanks Nahida for drafting assistance, benchmark execution,
figure and table preparation, independent validation against ExaLogLog,
and iterative review throughout the development of this manuscript. The
work conducted by the U.S. Department of Energy Joint Genome Institute
(https://ror.org/04xm1d337), a DOE Office of Science User Facility, is
supported by the Office of Science of the U.S. Department of Energy
operated under Contract No.~DE-AC02-05CH11231.

\textbf{Disclosure of AI assistance.} This work made extensive use of
large language model (LLM) AI tools, including Anthropic Claude and
Google Gemini. The initial draft of this manuscript was prepared with AI
assistance; Gemini provided detailed editorial feedback and suggestions.
All algorithmic design decisions, experimental methodology,
interpretation of results, and scientific conclusions are solely those
of the author.

\hypertarget{references}{%
\section*{References}\label{references}}
\addcontentsline{toc}{section}{References}

\hypertarget{refs}{}
\begin{cslreferences}
\leavevmode\hypertarget{ref-ertl2025exa}{}%
{[}1{]} O. Ertl, ``ExaLogLog: Space-efficient and practical approximate
distinct counting up to the exa-scale,'' \emph{arXiv preprint
arXiv:2402.13726}, 2024.

\leavevmode\hypertarget{ref-flajolet2007}{}%
{[}2{]} P. Flajolet, É. Fusy, O. Gandouet, and F. Meunier,
``HyperLogLog: The analysis of a near-optimal cardinality estimation
algorithm,'' \emph{Discrete Mathematics and Theoretical Computer
Science}, vol. AH, pp. 137--156, 2007.

\leavevmode\hypertarget{ref-bbtools}{}%
{[}3{]} B. Bushnell, ``BBTools: BBMap short-read aligner and other
bioinformatic tools.'' \url{https://sourceforge.net/projects/bbmap/},
2014.

\leavevmode\hypertarget{ref-flajolet1985}{}%
{[}4{]} P. Flajolet and G. N. Martin, ``Probabilistic counting
algorithms for data base applications,'' \emph{Journal of Computer and
System Sciences}, vol. 31, no. 2, pp. 182--209, 1985.

\leavevmode\hypertarget{ref-ertl2024}{}%
{[}5{]} O. Ertl, ``UltraLogLog: A practical and more space-efficient
alternative to HyperLogLog,'' \emph{Proceedings of the VLDB Endowment},
vol. 17, no. 7, pp. 1655--1668, 2024, doi:
\href{https://doi.org/10.14778/3654621.3654632}{10.14778/3654621.3654632}.

\leavevmode\hypertarget{ref-bushnell2026dll}{}%
{[}6{]} B. Bushnell, ``DynamicLogLog: Faster, smaller, and more accurate
cardinality estimation.'' 2026. Available:
\url{https://arxiv.org/abs/2603.27405}

\leavevmode\hypertarget{ref-karppa2022}{}%
{[}7{]} M. Karppa and R. Pagh, ``HyperLogLogLog: Cardinality estimation
with one log more,'' in \emph{Proceedings of the 28th acm sigkdd
conference on knowledge discovery and data mining}, 2022, pp. 753--763.
doi:
\href{https://doi.org/10.1145/3534678.3539246}{10.1145/3534678.3539246}.

\leavevmode\hypertarget{ref-wang1997}{}%
{[}8{]} T. Wang, ``Integer hash function.''
\url{https://web.archive.org/web/20071223173210/http://www.concentric.net/~Ttwang/tech/inthash.htm},
1997.
\end{cslreferences}

\end{document}